\documentclass[pre,twocolumn,nofootinbib,twoside,showpacs,superscriptaddress,tightenlines,floatfix]{revtex4-1}
\usepackage{dcolumn}
\usepackage{lipsum}
\usepackage{graphicx,amssymb,amsmath,epsf,bm}
\usepackage[utf8x]{inputenc}
\usepackage[T1]{fontenc}
\usepackage[normalem]{ulem}
\usepackage{color}
\usepackage{mathtools}
\usepackage{dsfont}
\usepackage{nicefrac}
\usepackage{hyperref}
\hypersetup{
    allcolors=blue,
    colorlinks=true,
    bookmarks=true,
    pdfpagemode=FullScreen,
}
\definecolor{nred}{RGB}{224,0,0}
\definecolor{nblue}  {RGB}{28,130,185}
\definecolor{dgreen} {RGB}{38,238,21}
\definecolor{norange}{RGB}{230,120,20}

\newcommand{\Tr}{{\rm Tr}}

\begin{document}


\title{$O(N)\times O(2)$ scalar models: including $\mathcal{O}(\partial^2)$ corrections
in the Functional Renormalization Group analysis.}

\author{Carlos A. S\'anchez-Villalobos}
\affiliation{Instituto de F\'isica, Facultad de Ciencias, Universidad de la Rep\'ublica, Igu\'a 4225, 11400, Montevideo, Uruguay}%
\affiliation{Sorbonne Universit\'e, CNRS, Laboratoire de Physique Th\'eorique de la Mati\`ere Condens\'ee, LPTMC, 75005 Paris, France}%

\author{Bertrand Delamotte}
\affiliation{Sorbonne Universit\'e, CNRS, Laboratoire de Physique Th\'eorique de la Mati\`ere Condens\'ee, LPTMC, 75005 Paris, France}%

\author{Nicol\'as Wschebor}
\affiliation{Instituto de F\'isica, Facultad de Ingenier\'ia, Universidad de la
Rep\'ublica, J.H.y Reissig 565, 11000 Montevideo, Uruguay
}%

\begin{abstract}
The study of phase transitions in frustrated magnetic systems with $O(N)\times O(2)$ symmetry has been the subject of controversy for more than twenty years, with theoretical, numerical and experimental results in disagreement. Even theoretical studies lead to different results, with some predicting a first-order phase transition while others find it to be second-order. Recently, a series of results from both numerical simulations and theoretical analyses, in particular those based on the Conformal Bootstrap, have rekindled interest in this controversy, especially as they are still not in agreement with each other. Studies based on the functional renormalization group have played a major role in this controversy in the past, and we revisit these studies, taking them a step further by adding non-trivial second order derivative terms to the derivative expansion of the effective action. We confirm the first-order nature of the phase transition for physical values of $N$, i.e. for $N=2$ and $N=3$ in agreement with the latest results obtained with the Conformal Bootstrap. We also study an other phase of the $O(N)\times O(2)$  models, called the sinusoidal phase, qualitatively confirming earlier perturbative results.
\end{abstract}

\maketitle

\section{Introduction}

The nature of the phase transition in scalar models with $O(N)\times O(2)$ symmetry has been the subject of a long-standing controversy for more than two decades \cite{Kawamura_1998,Pelissetto:2000ne,Calabrese:2002af,Calabrese:2003ib,Delamotte:2003dw,Calabrese:2004nt,Delamotte:2010ba,Kompaniets:2019xez}. These models are expected to describe two important families of frustrated magnets: the stacked triangular antiferromagnets (STAs) and the helimagnets. The above controversy is therefore not just theoretical but has experimental implications~\cite{Kawamura:1988zz,Delamotte:2003dw}.
In a Ginzburg-Landau formulation, such models can be described by means of a pair of $N-$component fields, $\vec \varphi_1$ and
$\vec \varphi_2$ and a Hamiltonian of the form:
\begin{align}
H&=\int d^dx \Big\{\frac 1 2 (\partial_\mu\vec\varphi_1)^2+\frac 1 2 (\partial_\mu\vec\varphi_2)^2
+\frac{r}{2}\big(\vec\varphi_1^2+\vec\varphi_2^2\big)\nonumber\\
&+\frac{u}{4!}\big(\vec\varphi_1^2+\vec\varphi_2^2\big)^2+\frac{v}{4!}\Big(\frac{1}{2}(\vec \varphi_1^2-\vec \varphi_2^2)^2+2(\vec\varphi_1\cdot\vec\varphi_2)^2\Big)\Big\}
\label{hamilitonian}
\end{align}
For $v>0$, the ground state corresponds to orthogonal $\vec \varphi_1$ and $\vec \varphi_2$ of the same modulus and is called orthogonal. As the temperature is varied, this system undergoes a phase transition to a 
paramagnetic phase and the controversy concerns the order of the transition for $N=2$ and $3$, which are the physically interesting cases.

\begin{figure}[t]
    \centering
    \includegraphics[width=8.8cm]{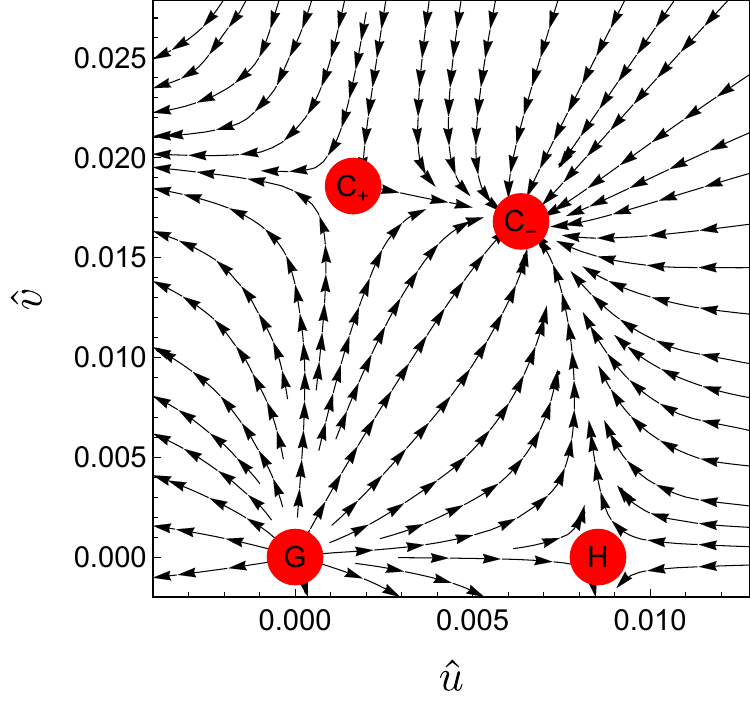}
        \includegraphics[width=8.8cm]{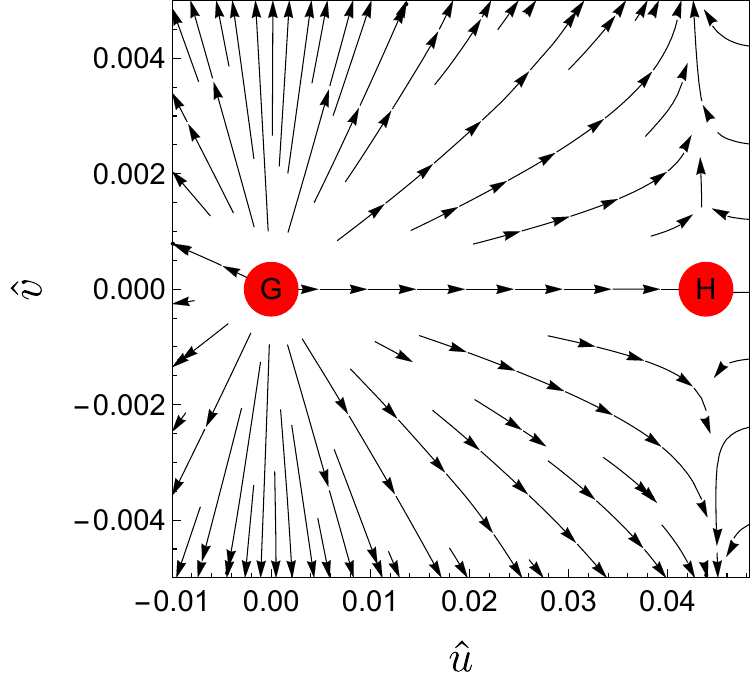}
    \includegraphics[width=8.8cm]{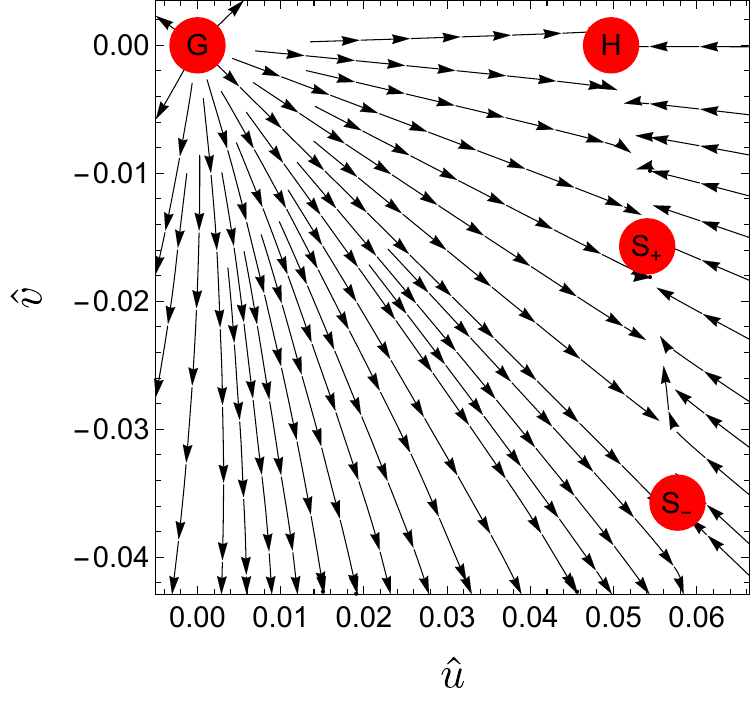}
        \includegraphics[width=8.8cm]{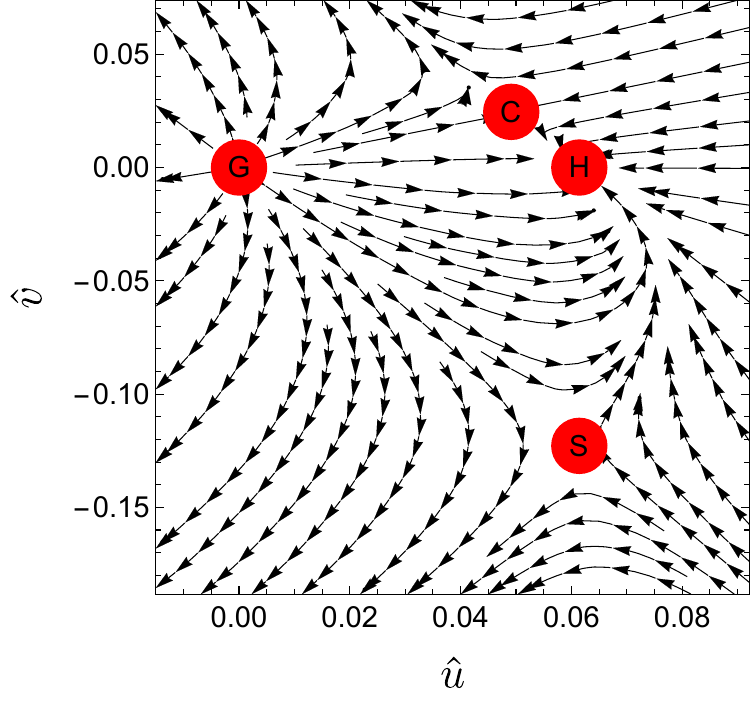}
\caption{Flows and perturbative fixed points in $d=3.9$ in the one-loop approximation for different values of $N$. Top left: $N=32>N_c^+$; Top right: $N_c^-<N=3<N_c^+$; Bottom left: $N_c^{H}<N=2.18<N_c^-$; Bottom right: $N=1<N_c^{H}$. Figure inspired from \cite{Henriksson:2020fqi}. The represented couplings are the dimensionless $\hat{u}= w_d \tilde{u}$ and $\hat{v}= w_d \tilde{v}$ where $w_d^{-1}=2^{d-1}\pi^{d/2}\Gamma(d/2)$ (see Eq.~(\ref{dimensionless})). The Gaussian fixed point is called G and the Heisenberg (or $O(2N)$ invariant) fixed point  H. For $N=1$, the chiral fixed point is denoted by $C$ and the sinusoidal by $S$. \label{fig:fixedPoint} }
\end{figure}

To present the problem, let us first recall what happens near the upper critical dimension $d_c=4$ of this system around which it is possible to perform a controlled perturbative expansion in $\epsilon=4-d$ \cite{Jones_1976,Bailin_1977,Kawamura:1988zz}. Even for  small $\epsilon$,  this  model has a rich structure of fixed points of the renormalization group (RG) which varies according to the value of $N$, see Fig.~\ref{fig:fixedPoint}:
\begin{itemize}
\item[i)] For $N>N_c^+(d)$, with
\begin{align}
 N_c^+(d)&=4(3+\sqrt{6})-4\Big(3+\frac{7}{\sqrt{6}}\Big)\epsilon+O(\epsilon^2)\nonumber\\
 &\simeq 21.8-23.4\epsilon+O(\epsilon^2),
\end{align}
the model has four perturbative fixed points. The first two are the Gaussian fixed point ($G$)
and the standard Heisenberg ($H$) or Wilson-Fisher $O(2N)-$invariant fixed point, corresponding both to multicritical transitions. In addition to these, it has two ``anisotropic'' fixed points usually referred to as ``chiral'' ($C_+$) and ``anti-chiral'' ($C_-$). The $C_+$ fixed point is once unstable and is associated with a phase transition between the ordered orthogonal phase and a paramagnetic phase. The $C_-$ fixed point is twice unstable and is associated with the tricritical transition between the  same two phases.

\item[ii)] For $N_c^+(d)>N>N_c^-(d)$, with
\begin{align}
 N_c^-(d)&=4(3-\sqrt{6})-4\Big(3-\frac{7}{\sqrt{6}}\Big)\epsilon+O(\epsilon^2)\nonumber\\
 &\simeq 2.2-0.57\epsilon+O(\epsilon^2),
\end{align}
only the Gaussian and Heisenberg fixed points exist. They are both multicritical and the system undergoes generically a first-order phase transition. By decreasing $N$ at fixed $d$, $C_+$ and $C_-$ approach each other, ``annihilate'' for $N=N_c^+(d)$ and become complex for $N_c^+(d)>N>N_c^-(d)$.
\item[iii)] For $N_c^-(d)>N>N_c^{H}(d)$, with
\begin{equation}
 N_c^{H}(d)=2-\epsilon+O(\epsilon^2),
\end{equation}
the model has again four fixed points but, unlike in case (i), in addition to the Gaussian and Heisenberg fixed points, it has two fixed points called ``sinusoidal'' ($S_+$) and ``anti-sinusoidal'' ($S_-$). These correspond to transitions between a paramagnetic phase
and an ordered phase where $\vec \varphi_1$ and $\vec\varphi_2$ are parallel in the ground state. The ``sinusoidal'' fixed point corresponds to a second-order phase transition and the others to multicritical transitions. As before, the line $N_c^-(d)$ corresponds to the location where  $S_+$ and $S_-$ ``annihilate''.
\item[iv)]For $N<N_c^{H}(d)$, the Heisenberg fixed point exchanges stability with the sinusoidal one that becomes a chiral fixed point $C$ and the second-order transition has an emergent $O(2N)-$symmetry at large distance. 
\end{itemize}

It is important to note that only cases (i) and (ii) are relevant for the  phase transition in the STA, but the other cases are of theoretical interest for a better understanding of the model, which is why this article deals with all of them.

\onecolumngrid

\twocolumngrid

The above results become controversial when the curve $N_c^+(d)$ is prolonged down to $d=3$ \cite{Kawamura_1998,Pelissetto:2000ne,Calabrese:2002af,Calabrese:2003ib,Delamotte:2003dw,Calabrese:2004nt,Delamotte:2010ba,Kompaniets:2019xez}. First of all, among the methods based on resummed perturbation theory, the  $\epsilon=4-d$ expansion at 6-loops presents a first order transition for $N=2,$ and $3$ \cite{Kompaniets:2019xez} but the perturbative expansion (in the constants $u$ and $v$) in fixed dimension presents a second order transition for those values of $N$. This expansion is performed at 5-loops in the $\overline{MS}-$scheme \cite{Calabrese:2004nt} and at 6-loops in the mass zero-momentum scheme \cite{Pelissetto:2000ne,Calabrese:2002af,Calabrese:2003ib}.

\begin{figure}[t]
    \centering
    \includegraphics[width=8.4cm]{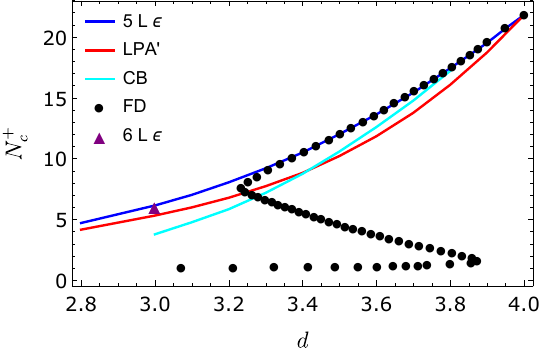}
    \caption{Figure based on Fig. 7 of Ref.~\cite{Delamotte:2010ba} showing previous results for $N_c^+(d)$. The blue curve shows the five-loop result of the $\epsilon$-expansion \cite{calabrese2004five}, the dots show the $\overline{MS}$ scheme results in fixed dimension \cite{Delamotte:2010ba}, the red curve shows the FRG results obtained with the LPA' \cite{Delamotte:2003dw}, the cyan curve shows the lower bound given by Conformal Bootstrap \cite{Reehorst:2024vyq}, the triangle shows the six-loop $\epsilon$-expansion result in $d=3$ \cite{Kompaniets:2019xez}. The part above the curve corresponds to a second order phase transition and the part below to a first order phase transition.}
    \label{fig:Ncd}
\end{figure}

In fact, as shown in Fig.~\ref{fig:Ncd}, the curves corresponding to both these perturbative analyses start almost identically near $d=4$ but the curve of the $\epsilon-$expansion is  monotonous while the curve given by the perturbative expansion at fixed dimension has a ``S''-shape around $d=3$ that puts the values $N=2$ and $N=3$ on the second order side of the $N_c^+(d)$ curve in $d=3$. It should be noted, as discussed in Ref.~\cite{Delamotte:2010ba}, that the shape of the ``S'' curve is quite unstable and depends substantially on the details of the resummation procedure employed. In Fig.~\ref{fig:Ncd}, an example is shown in which the return point occurs at $d \approx 3.2$ but in some other resummation procedures (see, for example, \cite{Calabrese:2004nt}) the return point occurs slightly below $d=3$ and the upper branch of the ``S'' curve looks very similar to the  $\epsilon-$expansion even in $d=3$. However, as $N$ is further decreased in $d=3$,
a second order zone including the points $N=2$ and $N=3$ is found in resummed results at fixed dimension. 

Other perturbative limits have also been studied including the $\epsilon=d-2$ expansion  ~\cite{Azaria_90} and the large$-N$ expansion \cite{Bailin_1977,Kawamura:1988zz,Pelissetto:2001fi,Gracey:2002ze,Gracey:2002pm} but the situation was not clarified by these results in $d=3$ for $N=2$ or $3$. It is worth mentioning that for $N\gtrsim 6$, the estimates of the critical exponents in $d=3$ obtained by the different methods (except the $d-2$ expansion) agree, so there is no controversy for these values of $N$.

The Functional Renormalization Group (FRG), also called the Non-Perturbative Renormalization Group, has  played a central role in the controversy described above \cite{Tissier:1999hv,Tissier:2000tz,Tissier:2001uk,Delamotte:2003dw,Debelhoir:2016yfp,yabunaka2016local}. These studies used a different approximation scheme from perturbation theory, called ``Derivative Expansion'' (DE)  (for a recent review, see Ref.~\cite{Dupuis:2020fhh}). Up to now, for $O(N)\times O(2)$ models, only the leading order of the DE (called Local Potential Approximation, LPA) or its variants (LPA') have been implemented.

The first studies in this problem using the LPA were carried out in the framework of the Wilson-Polchinski equation \cite{Zumbach_93} obtaining $N_c^+(d=3)\sim 4.7$. 
The topic made an important progress by the use of the Wetterich equation \cite{Wetterich:1992yh,Ellwanger:1993kk,Morris:1993qb,
Delamotte:2007pf,Dupuis:2020fhh}. It allowed to implement the LPA' (an improved version of the LPA) \cite{Tissier:1999hv,Tissier:2000tz,Tissier:2001uk}, as reviewed in Ref.~\cite{Delamotte:2003dw}. LPA' allowed to recover all perturbative limits  for $d$ near four and two, as well as in the large$-N$ limit. 
The corresponding $N_c^+(d)$ curve is shown in Fig.~\ref{fig:Ncd}. It exhibits a monotonous behavior and gives $N_c^+(d=3)=5.1$ which a priori implies a first order transition in $d=3$ for the physically interesting cases $N=2$ and $N=3$. Moreover, the critical exponents for $N\gtrsim 6$ are in good agreement with those of perturbation theory. The aforementioned analysis in the LPA' was carried out by expanding the effective potential around its minimum in powers of the two invariants present in this model. This field-expansion seems to converge very well in dimensions $d\gtrsim 3$ and leads to results consistent with those of subsequent studies in which an expansion in only one of the invariants was performed \cite{Debelhoir:2016yfp}, or in which the complete function in both invariants was taken into account \cite{yabunaka2016local,Yabunaka:2024tmc}. It was shown in Ref.~\cite{Debelhoir:2016yfp} that even if the transition is first order, the correlation length at the transition is very large for a wide range of parameters of the Hamiltonian, that is, the transition is weakly first order. This explains why for these systems it is difficult from Monte Carlo simulations and from experiments (see below) to differentiate between a first-order transition and a second-order transition.
 
In many cases, controversies between different theoretical methods based on field theory have been elucidated by resorting to Monte-Carlo simulations and/or experimental results. Frustratingly enough, in this problem this has not been the case. In Refs.~\cite{Kawamura_1998,Delamotte:2003dw} a review of classical experimental and numerical results is given.  As before, Monte-Carlo results are in agreement with all other methods for $N\gtrsim6$ in $d=3$. It should be noted that for these values of $N$, Monte Carlo results can be used as a benchmark for various approximate methods. 
For $N=2$ and $3$, power laws are observed for most quantities studied which suggests a second order transition.
At the same time, they show violations of universality, violations in the scaling relations between exponents and results incompatible with reflection positivity, such as negative anomalous dimensions of the field (for a detailed discussion, see Ref.~\cite{Delamotte:2003dw}). While it is true that reflection positivity is not a first-principle requirement for a statistical mechanical system, in practice the models used, such as the one in Eq.~(\ref{hamilitonian}), exhibit such a property and we can therefore expect at least the dominant behavior near criticality to respect this property.~\footnote{As mentioned below, models such as Eq.~(\ref{hamilitonian}) should give not only a dominant behavior compatible with unitarity but also scaling corrections compatible with it. However, it is important to keep in mind that there are possible scaling corrections in other models in the same universality class (such as those including terms corresponding to next to nearest neighbor interactions) that, in general, do not respect reflection positivity.}
To explain all the results above, it was proposed in Ref.~\cite{Delamotte:2003dw} that for $N\lesssim 6$ the transition is weakly first order, thus explaining the power law behaviors observed over a wide range of control parameters (such as temperature) and violations of universality and of scaling relations among exponents. It should be noted that more recent simulations have not clarified the overall picture as, despite the increased computational power, some seem to indicate a first-order transition \cite{Ngo_08,Ngo_08bis,Sorokin:2014whz,Sorokin:2022zwh} and one a second-order transition \cite{Nagano:2019lce}.

In this long-standing and as yet unresolved controversy, the Conformal Bootstrap method has recently come to the fore \cite{Henriksson:2020fqi,Reehorst:2024vyq}. At least in principle, this method makes it possible to obtain rigorous bounds on exponents and other critical properties. They are based on conformal invariance for  theories satisfying reflection positivity (for a review of the subject, see \cite{Poland:2018epd}). Such bounds have been shown to be stringent in several cases
\cite{Kos:2014bka,Kos:2016ysd,Chester:2019ifh}, allowing to reach, for example, the best estimate of the critical exponents to date for the three-dimensional Ising universality class \cite{Kos:2014bka}.
In practice, in order to reach precision exponents, on top of properties which in principle could be rigorously invoked such as conformal invariance or reflection-positivity, complementary hypotheses on the spectrum of the scaling operators of a critical point are added. These are often reasonable but they cannot always be justified on the basis of first principles. In many cases the results, at least for the dominant exponents or for the mere existence of a conformal theory, do not depend on these complementary hypotheses, but in others they may depend on them. This explains how a method that gives a priori rigorous bounds can sometimes obtain results that are not entirely certain. 

Such complementary hypotheses explain the possible disagreement between Conformal Bootstrap works on $O(N)\times O(2)$ models. In Refs.~\cite{Nakayama:2014sba,Henriksson:2020fqi} a critical point candidate with symmetry $O(N)\times O(2)$ is found for $N=2$ and $N=3$ (although the authors of Ref.~\cite{Henriksson:2020fqi} are extremely cautious about its existence). In apparent contradiction with this result, a lower bound for $N_c^+(d)$ has been found recently in~\cite{Reehorst:2024vyq} for all dimensions between $3$ and $4$, giving $N_c^+(d=3)>3.78$. 
The curve obtained in this reference for the lower bound of $N_c^+(d)$ is qualitatively similar to the curve obtained by the $\epsilon-$expansion or the FRG and does not show any indication down to $d=3$ of the existence of an ``S'' behavior. In this sense, this work suggests the existence of a first-order phase transition for $N=2$ and $N=3$ in $d=3$.

As pointed out in \cite{Reehorst:2024vyq}, it is possible that the Conformal Bootstrap results are not contradictory because
 neither the ``S''-shape of the curve $N_c^+(d)$ is ruled out in~\cite{Reehorst:2024vyq} nor the possibility of a region of second order phase transitions under the form of an ``island'' disconnected from the region above the curve $N_c^+(d)$. 
 On the other hand, as pointed out in this reference, the Conformal Bootstrap as well as the perturbative calculations and the Monte Carlo simulations \cite{Nagano:2019lce} are based on models that satisfy reflection positivity. As mentioned above, this implies that all exponents, including correction to scaling exponents, must be real which rules out the existence of a focus fixed point (focus fixed points have complex correction to scaling exponents that causes the renormalization flow to converge to them in a spiral fashion). However, both fixed dimension perturbation theory \cite{Pelissetto:2000ne,Calabrese:2002af,Calabrese:2003ib} and some Monte Carlo simulations \cite{Nagano:2019lce} seem to find focus fixed points.
 In this sense, the authors of Ref.~\cite{Reehorst:2024vyq} tend to suggest an apparent disagreement with Ref.~\cite{Henriksson:2020fqi} that could be due to some differences in the complementary assumptions about the spectrum of operators between these two works.

It should be noted, however, that the results from the FRG, which have played a central role in establishing the problem and defending one of the points of view in the controversy, had not incorporated until recently notable advances that have been made in this methodology in recent years. In a recent preprint \cite{Yabunaka:2024tmc} some of these methodological progress have been included at the level of LPA' but the inclusion of second order functions in the ansatz has been missing up to now. This is the main goal of this paper. With this in mind, it is worth briefly summarizing the methodological advances made in the DE of the FRG in recent years.

The DE consists, on the one hand, in regularizing the long-wavelength fluctuations at a momentum scale $k$ and, on the other, in projecting the Gibbs free energy onto a functional form comprising terms with up to a given number of derivatives of the field (see Sec.~\ref{sec_DE} for more details). It has recently been realized that this expansion has a small parameter, of order $1/4$, that characterizes the order of magnitude of successive terms that, consequently show an apparent convergence \cite{Balog:2019rrg,DePolsi2020,DePolsi2021,DePolsi2022}.
However, it has also been proven in those references that the existence of this small parameter depends crucially on the appropriate choice of the infrared regulator profile. 
Without going into technical aspects, the overall scale of such a regulator needs to be optimized \cite{Canet2003,Balog:2019rrg} and this has been done during decades using the Principle of Minimal Sensitivity (PMS) originally formulated in the framework of perturbation theory \cite{Stevenson:1981vj}. More recently it has been proven that, near a fixed point, this criterion is essentially equivalent to minimizing the breaking of conformal invariance at each order of the DE \cite{Balog2020,Delamotte:2024xhn}.
Another significant advance made in recent years has been the implementation at LPA \cite{Leonard15,yabunaka2016local,Yabunaka:2024tmc} and beyond \cite{Chlebicki:2022pxm} of fully functional approximations when more than one invariant is allowed by symmetries.

It should be noted that none of these technical advances have been applied to the study of frustrated magnetic systems beyond the leading order (LPA or LPA'). However, the inclusion of the second order terms of the DE is crucial to calibrate the quality of the approximation and to be able to estimate error bars \cite{DePolsi2020,DePolsi2021,Chlebicki:2022pxm}. This is done for the first time in this paper for $O(N)\times O(2)$ models.
As will be explained in detail below, the inclusion of all possible terms in the second order of the DE is a formidable calculation that is not feasible to achieve in the present work. Instead, we will limit ourselves to include the dominant terms of that order with the goal of studying their impact on the results obtained previously in the dominant order.

Our main result, explained below, is to  confirm the LPA (and LPA') prediction of the existence of a weakly first order transition for $N=2$ and $N=3$ in $d=3$. In particular, we show that the $N_c^+(d)$ curve can be monotonically extended down to $d=2.4$ without exhibiting any ``S'' type behavior. Below this dimension our numerical code becomes very demanding
and it is not possible to lower $d$ any further. However, our curve $N_c^+(d)$ is not only monotonous but even convex so very far from any indication of an ``S'' behavior even in very low dimensions.~\footnote{It is worth mentioning that in a less precise approximation \cite{yabunaka2016local} the dimension $d=2.3$ has been previously reached in the FRG framework without any indication of an ``S'' behavior.} These approximate results do not have the same level of rigor as the Conformal Bootstrap \cite{Henriksson:2020fqi,Reehorst:2024vyq}. However, they are complementary because we are able to compute $N_c^+(d)$ down to $d=2.4$. Furthermore, by integrating the RG flow in $d=3$ from ``natural'' initial conditions and after fine-tuning of the parameter $r$ in Eq.~\eqref{hamilitonian}, we easily find the $C_+$ fixed points for $N \gtrsim 6$, but not for $N=2$ and 3. Although this is not a definitive proof of the non-existence of this fixed point for these values of $N$, it does show that if it exists, the Hamiltonian \eqref{hamilitonian}  for $N=2$ and 3 with natural coupling constants does not belong to its basin of attraction. These results are a significant contribution to the controversy discussed above.

In addition to the analysis of the $N_c^+(d)$ controversy, we also calculate $N_c^-(d)$ and $N_c^H(d)$ which have not been studied with FRG, probably because they are not directly relevant to frustrated systems. The aim of this study is to test the quality of our approximations in other parameter regions. 
We also determine the form of the Gibbs free energy in the large-$N$ limit, thus generalizing to $O(N)\times O(2)$ a known result in $O(N)$ models \cite{DAttanasio:1997yph}. This allows us to prove that here again the effective potential at LPA is exact for $N\to \infty$.

The plan of the article is the following. First, we present in Sect.~\ref{secFRG} a succinct way and with the purpose of making the article self-contained, the main ideas of the Functional Renormalization Group and of the approximation scheme employed in this work (Derivative Expansion). After that, in Sect.~\ref{sec_Results} the results are presented. We start by explaining the precise way in which the error bars are calculated.
Then, we present the results, including the $N_c^+(d)$, $N_c^-(d)$ and $N_c^H(d)$ curves and the exponents obtained for $d=3$ and  $N \ge 6$.  Moreover, a study carried out directly for $d=3$ and $N=2$ and $3$ is presented with the purpose to search for a critical fixed point for these cases. Finally, conclusions and perspectives of the present work are presented. Material of a more technical nature is presented in appendices or in the supplementary material. 

\section{The functional renormalization group and the Derivative Expansion}\label{secFRG}

In this section we make a brief reminder of the fundamental aspects of the FRG and of the approximation scheme we used in this work: the Derivative Expansion.
The presentation of this method has been done in detail on multiple occasions, so the presentation here will be done in a very succinct manner in order to introduce the basic notations to be used later. For a general discussion of the FRG and the Derivative Expansion, see \cite{Dupuis:2020fhh,Delamotte:2007pf}; for more details on the convergence properties of the derivative expansion and the calculation of error bars, see \cite{Balog2019,DePolsi2020};
for a review on its implementation in the context of frustrated magnets, see \cite{Delamotte:2003dw}.

\subsection{The FRG in a nutshell}

The FRG is a modern version of the Wilson (or Wilson-Polchinski) Renormalization Group. An exact equation determining its renormalization flow is known: the Wetterich equation \cite{Wetterich:1992yh,Ellwanger:1993kk,Morris:1993qb}. The FRG is strictly equivalent to the Wilson-Polchinski Renormalization Group \cite{Polchinski:1983gv} from a formal point of view, one being obtained from the other by means of a Legendre transform (see, for example, \cite{Morris:1993qb}). For reasons explained below, we will only use the scheme based on the scale-dependent effective action.

Let us consider a theory described by $\mathcal{N}$ real scalar fields regularized in the infrared by adding the following term to its Hamiltonian \cite{Polchinski:1983gv}:
\begin{equation}\label{deltaHDirect}
    \Delta H_k[\varphi]=\frac 1 2 \int_{x,y}\varphi_i(x) R_k(x-y) \varphi_i(y),
\end{equation}
where $\int_x=\int d^dx$. Here and in the rest of the article, we employ Einstein's convention unless stated otherwise. We choose $\Delta H_k$ $O(\mathcal{N})$-invariant even if the model has a lower symmetry.
For dimensional reasons one usually writes the regulator profile in Fourier space as: \footnote{We employ the same symbol for the regulator profile $R_k$ and its Fourier transform. Similarly, we use the same criterion for its dimensionless counterparts.}
\begin{equation}\label{eq:RegProf}
    R_k(q^2)=\alpha Z_k k^2 r(q^2/k^2).
\end{equation}
In this expression we have factored-out a global scale of the regulator, $\alpha$,  and a field renormalization factor, $Z_k$, whose determination will be specified later (see Eq.~(\ref{eq:defZk})).
For $R_k$ to be an efficient infrared regulator, the function
$r(q^2/k^2)$ must decrease faster than any power of $q^2$ at large momentum and tend to one for $q\ll k$. Note, however, that regulators that diverge at $q\to 0$ have also been studied.
In the present work we will always employ the Wetterich regulator:
\begin{equation}
 R_k(q^2)=\alpha Z_k \frac{q^2}{e^{q^2/k^2}-1},
\end{equation}
that has shown in the past to give excellent results
once the value of $\alpha$ has been optimized
via the PMS procedure (see, for example, \cite{DePolsi:2020pjk}).

The regularized partion function in the presence of a source for the $\varphi_i(x)$ fields reads:
\begin{equation}\label{functIntegral}
	Z_k[J]=\int \mathcal{D}\varphi e^{-H[\varphi]-\Delta H_k[\varphi]+\int_x J_i(x)\varphi_i(x)}.
\end{equation}
It will be implicitly assumed all along this article that the theory has been regularized somehow in the ultraviolet and we call $\Lambda$ the scale of the ultraviolet cut-off.
$W_k[J]=\log Z_k[J]$ is the Helmholtz free-energy or generating functional
of connected correlation functions. The Gibbs free-energy, or
scale-dependent effective action is defined as the modified Legendre
transform of $W_k[J]$:
\begin{equation}
 \Gamma_k[\vec \phi]=\int_x \phi_i(x) J_i(x) -W_k[J]-\Delta H_k[\vec \phi].
\end{equation}
where $\vec \phi(x)=\big(\phi_1,\dots,\phi_{\mathcal{N}}\big)$. 
In the previous equation, $J_i(x)$ is an implicit function of the $\phi_i(x)$,
obtained by inverting
\begin{equation}
  \phi_i(x)=\frac{\delta W_k}{\delta J_i(x)}.
\end{equation}

$\Gamma_k[\vec \phi]$ is the generating functional of IR-regularized proper vertices or amputated 1PI correlation functions (see \cite{Dupuis:2020fhh} for details). 
This is at odds with the Wilson-Polchinski functional which is the generating
functional of connected diagrams with amputated external legs. This leads to the fact that the DE expansion parameter is of order one when the expansion is performed in the Polchinski equation but is of order $1/4$ in the Wetterich equation \cite{Balog2019,DePolsi2020}. 

In the rest of the article, we will omit the $k$-dependence of the regularized proper vertices to alleviate notation. In
actual calculations, we will be interested in proper vertices
evaluated in a uniform field which only depend on $n-1$ independent wave-vectors due to translation invariance.

The exact evolution of $\Gamma_k[\vec \phi]$ is given by Wetterich equation \cite{Wetterich:1992yh,Ellwanger:1993kk,Morris:1993qb} (below $t=\log(k/\Lambda)$):  
\begin{equation}
\partial_{t}\Gamma_{k}[\vec\phi]=\frac{1}{2}\int_{x,y}\partial_{t}R_{k}(x-y)G_{ii}[x,y;\vec \phi]
\label{wettericheq}
\end{equation}
where $G_{ij}[x,y;\vec \phi]$ is the full propagator in an arbitrary external
field $\vec \phi(x)$. The full propagator is the inverse
of the two-point vertex (adding to it the regulator function).

From Eq.~(\ref{wettericheq}) follows the exact flow equation of the effective potential $U_k$ defined, up to a volume factor, by $\Gamma_k[\phi]$ evaluated in a constant field:
\begin{equation}
\label{eqpot}
 \partial_t U_k(\vec\phi)=\frac 1 2 \int_q \partial_t R_k(q) G_{ii}(q;\vec\phi).
\end{equation}
where $G_{ij}(q;\vec\phi)$ is the Fourier transform of the full propagator evaluated in a uniform field. Furthermore, one can derive flow equations for all vertex functions evaluated in a constant field by functionally derivating
Eq.~(\ref{wettericheq}) and then evaluating in a uniform field.
For instance, for the 2-point function in a uniform external field it corresponds to:
\begin{align}
 \partial_t &\Gamma_{ij}^{(2)}(p;\phi)=\int_q \partial_t R_k(q) G_{mn}(q;\phi)
 \Big\{-\frac 1 2 \Gamma_{ijns}^{(4)}(p,-p,q;\phi)\nonumber\\
&+ \Gamma_{inl}^{(3)}(p,q;\phi)G_{lr}(p+q;\phi)\Gamma_{jsr}^{(3)}(-p,-q;\phi)
 \Big\}G_{sm}(q;\phi)
 \label{eqvertex2}
\end{align}
As is
well-known, this is the beginning of an infinite hierarchy of coupled equations,
where the equation for $\Gamma^{(n)}$ depends on all the vertices up
to $\Gamma^{(n+2)}$.  This infinite hierarchy for vertex functions cannot be
solved without approximations, in most cases.

The main advantage of the FRG with respect to other field-theoretical methods is that it provides a robust framework for various types of ``non-perturbative'' approximations. As announced in the introduction, in this paper we will use the Derivative Expansion approximation scheme. In the following section we proceed to summarize its main ideas. 
 
\subsection{The derivative expansion: the general case}
\label{sec_DE}
The most widely employed approximation scheme in the framework of the FRG is the DE. It consists, at order $\mathcal{O}(\partial^s)$, in truncating the momentum-dependence of the vertices up to the power $s$.  This approximation applies when one is interested in the long-distance properties of the theory. In this case, FRG equations allows the controlled expansion of the regulated vertices at small momenta. In fact, it has been shown that this is an accurate and precise approximation scheme, for example, for $\mathbb{Z}_2$, $O(N)$-invariant and $\mathbb{Z}_4$ models (see for example, \cite{Canet2003b,Canet2003,Balog2019,DePolsi2020,DePolsi2021,Chlebicki:2022pxm}). Moreover, it has been proven to be an extremely robust approximation in a large variety of problems.
To give just a few examples, it has been applied with great success to the study of the  Random Field Ising model (spontaneous supersymmetry breaking and the associated breaking of dimensional 
reduction in a nontrivial dimension) \cite{Tarjus:2004wyx,Tissier:2011zz},
the glassy phase of crystalline membranes \cite{Coquand2017},  systems showing different critical exponents in their high and low temperature phases \cite{Leonard15}, the phase diagram of reaction-diffusion systems \cite{Canet:2004je,Canet:2003yu}, the calculation of non-universal critical temperatures of spin systems \cite{Machado:2010wi}. For a large variety of other examples, see \cite{Dupuis:2020fhh}.

In this section, we present such an approximation in general up to order $\mathcal{O}(\partial^2)$ for a theory described by $\mathcal{N}$ real scalar fields leaving for the next section the $O(N)\times O(2)$ case.

Let us consider first the leading order of the approximation scheme, usually called the Local Potential Approximation (LPA) (or order $\mathcal{O}(\partial^0)$).  This order consists of projecting $\Gamma_k$ onto an {\it ansatz} including a scale-dependent effective potential and a bare kinetic term:
\begin{equation}\label{eq:LPA}
	\Gamma_k[\vec \phi]=\int_x \biggl\{\frac{1}{2}\partial_{\mu}\phi_i\partial_{\mu}\phi_i+U_k(\vec \phi)\biggr\},
\end{equation}
with $U_k(\vec \phi)$ the effective potential that must be invariant under the symmetry group of the considered problem.
This approximation corresponds to replacing all vertices by their zero-momenta value, except for the two-point vertex for which the bare dependence is maintained.

At $\mathcal{O}(\partial^2)$, $\Gamma_k[\vec \phi]$ is approximated by:
\begin{equation}\label{eq:DE2}
\Gamma_k[\vec\phi]=\int_x \biggl\{U_k(\vec\phi)+\frac{(Z_k)_{ij}(\vec\phi)}{2}\partial_{\mu}\phi_i\partial_{\mu}\phi_j\biggr\},
\end{equation}
where $(Z_k)_{ij}(\vec\phi)$ is a symmetric rank-two tensor. For example, in the $O(\mathcal{N})$ universality class, the {\it ansatz} takes the form:

\begin{equation}\label{eq:DE2ON}
\Gamma_k^{O(\mathcal{N})}[\vec\phi]=\int_x \biggl\{U_k(\rho)+\frac{Z_k(\rho)}{2}\partial_{\mu}\phi_i\partial_{\mu}\phi_i+\frac{Y_k(\rho)}{4}\partial_{\mu}\rho \partial_{\mu}\rho\biggr\},
\end{equation}
where $\rho=\phi_i\phi_i/2$ is the $O(\mathcal{N})-$invariant scalar and $U_k$, $Z_k$ and $Y_k$ are scalar functions.

The flow equation of the functions $(Z_k)_{ij}(\vec\phi)$ can be obtained 
by differentiating Eq.~(\ref{eqvertex2}) with respect to $p^2$ and evaluating at zero momentum. The flow equations for $Z_k$ and $Y_k$ are obtained by inserting {\it ansatz}~(\ref{eq:DE2ON}) into Eq.~(\ref{eqvertex2}).
 In a similar way, the expansion can be pushed to higher orders by including terms with higher number of derivatives. This has been done up to order $\mathcal{O}(\partial^6)$ in the Ising model \cite{Balog2019} and  to order $\mathcal{O}(\partial^4)$ in $O(N)$ models \cite{DePolsi2020,DePolsi2021}.

The reason for the success of the DE relies on the fact that its successive orders are suppressed by a factor of $\lambda\simeq 1/4$ \cite{Balog2019}. This estimate is confirmed by an empirical analysis of the Ising and $O(\mathcal{N})$ universality classes \cite{Balog2019,DePolsi2020,DePolsi2021}.
More recently, this analysis was improved by taking into account the dependence of $\lambda$ on the regulator profile \cite{DePolsi2022}. It has been shown in this reference that the observed convergence of the DE \cite{Balog2019,DePolsi2020,DePolsi2021} crucially depends on the tuning of the overall scale $\alpha$ of the regulator  in Eq.~\eqref{eq:RegProf} which is achieved using the Principle of Minimal Sensitivity (PMS). The PMS states that since any physical quantity should be independent of the shape of the regulator, a spurious dependence on this shape, induced by the truncation of the DE at a finite order, should be minimized. Accordingly, the optimal choice for $\alpha$ corresponds to $\alpha_{\rm opt}$ such that the quantity under study is stationary for $\alpha=\alpha_{\rm opt}$.

It should be noted that for the study of critical phenomena, it is convenient to dedimensionalize all variables and that it is only in terms of dimensionless variables that the Renormalization Group equations have fixed points. The passage to dimensionless variables is standard and we mention it here only for the purpose of fixing notations:
\begin{align}
\label{dimensionless}
	\tilde{\phi}_i &= w_d^{-1} Z_k^{\frac{1}{2}} k^{\frac{-d+2}{2}} \phi_i\nonumber\\
	\tilde{F}_k(\tilde{\vec{\phi}}) &=w_d^{-1+\frac{f}{2}} Z_k^{-\frac{f}{2}} k^{-d+\frac{f*(d-2)}{2}+ 2s} F_k(\vec{\phi})\nonumber\\
	\tilde{g}_k &=  w_d^{-1+\frac{f}{2}}Z_k^{-f/2} k^{-d+f\frac{(d-2)}{2}}g_k
\end{align}
where $w_d^{-1}=2^{d-1}\pi^{d/2}\Gamma(d/2)$, $f$ is the number of fields in the effective action ansatz multiplied to the function $F_k$ or the coupling constant $g_k$ that is included in the effective potential, and $s$ is the number of spatial derivatives applied to such fields.

The renormalization factor $Z_k$ introduced in the regulator (\ref{eq:RegProf}) and in the definition of dimensionless variables (\ref{dimensionless}) must be fixed in an appropriate way. In the present manuscript we employ a renormalization condition at zero external field:
\begin{equation}
\label{eq:defZk}
 Z_k=\frac{1}{\mathcal{N}}(Z_k)_{ii}(\vec\phi=\vec 0).
\end{equation}
Before proceeding to the next section, it is worth noting that
the DE equations for marginal and relevant couplings 
are exact to order $\epsilon$ for $d=4-\epsilon$.
Moreover, for $O(\mathcal{N})$ models, the flow equation of the effective potential is exact already at LPA for $\mathcal{N}\to \infty$.

\subsection{The derivative expansion: the $O(N)\times O(2)$ case}
\label{sec_DE_ONO2}

In this section we apply the DE, whose general scheme was presented in the previous section, to the case of a system, such as the one described by the Hamiltonian (\ref{hamilitonian}).

\subsubsection{LPA}
\label{sec:LPA}

The LPA ansatz for $\Gamma_k$ involves a kinetic term which is $O(2N)$-invariant, and thus also $O(N)\times O(2)$-invariant, and a potential $U_k$ which is a $O(N)\times O(2)$ scalar. By definition, such a scalar function is a function of the invariants built out of the fields and that does not involve any derivative.

All $O(N)$-invariant terms built from constant fields $\phi_1$ and $\phi_2$ must be functions of \begin{equation}
 S_{ab}=\vec \phi_a \cdot \vec \phi_b.
\end{equation}
As for the $O(2)$ invariance, all $O(2)$ indices must be contracted. That is, the most general $O(N)\times O(2)-$invariant potential can be written in terms of the invariants
\begin{equation}
 \tau_n= \Tr \Big(S^n\Big)=\frac 1 2 S_{a_1a_2}S_{a_2a_3}\cdots S_{a_{n}a_1}
\end{equation}

However, the $\tau_n$ invariants are not independent thanks to the Cayley-Hamilton theorem which in the present case implies that
\begin{equation}
 S^2-S \Tr(S)+\det(S) Id=0.
\end{equation}
This expression implies that all the $\tau_n$ for $n\geq 3$ can be expressed algebraically in terms of $\tau_1$ and $\tau_2$. In other words, the most general effective potential is a smooth function of the two invariants:
\begin{align}
 \tau_1&=\frac 1 2 \Tr \Big(S\Big)=\frac 1 2 \Big(\vec \phi_1^2+ \vec \phi_2^2\Big)=\rho,\nonumber\\
\tau_2&=\frac 1 2 \Tr \Big(S^2\Big)=\frac 1 2 \Big((\vec \phi_1^2)^2+ (\vec \phi_2^2)^2\Big)+\big(\vec \phi_1\cdot \vec \phi_1\big)^2.
\end{align}
Contrary to $\tau_1=\rho$, $\tau_2$ is not $O(2N)$-invariant. We call it anisotropic for this reason. We find convenient in the following to work with:
\begin{equation}
 \tau=2\tau_2-2\rho^2=\frac 1 2 \big(\vec \phi_1^2-\vec \phi_2^2\big)^2+2\big(\vec \phi_1\cdot \vec \phi_2\big)^2
\end{equation}
instead of $\tau_2$. Notice that the Hamiltonian (\ref{hamilitonian}) has already been expressed in terms of these two invariants.

For numerical purposes, the choice of $\rho$ and $\tau$ as independent variables is not convenient because $2\rho^2\ge \tau$ which makes non-trivial the domain  of admissible values of $\rho$ and $\tau$. Going back to the constant fields $\vec\phi_i$, one can show that using $O(N)\times O(2)$ transformations, it is possible to transform them in:
\begin{align}
 \vec \phi_1&=(\varphi_1,0,\dots,0),\nonumber\\
  \vec \phi_2&=(0,\varphi_2,0,\dots,0).
  \label{uniffield}
\end{align}
This allows working with a regular grid with coordinates $(\varphi_1,\varphi_2)$.
As was shown in Ref.~\cite{yabunaka2016local} this procedure works well in $d=3$ but deteriorates rapidly for lower dimensions in LPA due to the appearance of spurious eigenperturbations around the fixed point of the Renormalization Group when $d<3$.

To avoid this problem we adopt the procedure employed in Ref.~\cite{Chlebicki:2022pxm}.
In that case, instead of using the variables $(\varphi_1,\varphi_2)$, the following change of variables was performed:
\begin{align}
\label{rhoandtheta}
 \varphi_1&=\sqrt{2\rho} \cos \theta, \nonumber\\
 \varphi_2&=\sqrt{2\rho} \sin \theta.
\end{align}
In terms of the variables $(\rho,\theta)$ the boundary conditions are also trivial ($\rho \geq 0$ and $0 \leq \theta < 2 \pi$), but now the $O(2N)$ limit is reached exactly even within a finite grid (in the $O(2N)-$invariant limit the potential simply ceases to depend on $\theta$). As will be seen in Sec.~\ref{sec_Results} this set of variables avoids the numerical difficulties encountered in Ref.~\cite{yabunaka2016local} for low dimensions. 

After reducing the problem to the variables $\varphi_1$ and $\varphi_2$, a residual symmetry $\mathbb{Z}_4$ persists corresponding to the transformations
$\varphi_a \to -\varphi_a$ (with $a=1,2$) and exchange of $\varphi_1$ and $\varphi_2$. This residual symmetry allows the restriction to the region $\varphi_1\geq\varphi_2\geq 0$.  This $\mathbb{Z}_4$ invariance is expressed in variables $(\rho,\theta)$ by the periodicity of the potential in $\theta$ with a period $\pi/2$ on top of the invariance $\theta\to -\theta$. This allows to restrict $\theta$ to the interval $0 \leq \theta < \pi/4$.

The LPA, as discussed above, allows to recover exactly several limits. Firstly, it allows to obtain the exponents at order $\epsilon$ in $d=4-\epsilon$. 
Secondly, it has been observed that it also allows to obtain exactly the exponents in every dimension in the limit $N\to \infty$. 
This second point is well known for the $O(N)$ models \cite{DAttanasio:1997yph} and has been empirically observed for the $O(N)\times O(2)$ models previously \cite{Tissier:1999hv,Delamotte:2003dw} but no proof existed so far. In Appendix~\ref{app:largeN}, it is shown that the exactness of the equation for the effective potential and for the anomalous dimension in the $N\to \infty$ limit can also be generalized to the $O(N)\times O(2)$ models. 

As far as the behavior near the lower critical dimension $d=2+\epsilon$ is concerned, it has been shown that the inclusion of some scale-dependent {\it constants} in the terms with two derivatives (i.e., what is nowadays usually called LPA') allows to recover the behavior to order $\epsilon$. In the present work we will not include those contributions, that have been shown to have a very small role near $d=3$ (see Ref.~\cite{yabunaka2016local}).

\subsubsection{The second order of the Derivative Expansion}
\label{DE2}

We include in the present study some of the functions that appear at the second order of the DE that we call $\mathcal{O}(\partial^2)$ or DE2. In this section we detail that order in general and specify, in particular, the
specific terms included at  $\mathcal{O}(\partial^2)$ in the present work. The determination of the set of functions allowed by the $O(N)\times O(2)$ symmetry at order $\mathcal{O}(\partial^2)$ of the DE is not entirely trivial \footnote{In fact, we have found that the set proposed in Ref.~\cite{Delamotte:2003dw} omits several of the functions listed here; this omission is of no consequence in the above-mentioned reference but is important for the present work.} and is detailed in Appendix~\ref{apn:Ansatzes-DE2}. We find:
\begin{align}\label{eq:DE2ONO2}
&\Gamma_k^{O(N)\times O(2)}[\vec\phi]=\int_x \biggl\{U_k(\rho,\tau)+\frac{Z_k(\rho,\tau)}{2}\big(\partial_{\mu}\vec \phi_i\cdot \partial_{\mu}\vec \phi_i\big)\nonumber\\
&+\frac{Y_k(\rho,\tau)}{4}\partial_{\mu}\rho \partial_{\mu}\rho+
\frac{W^{(1)}(\rho,\tau)}{4}\big(\vec{\phi_1}\cdot\partial_\mu\vec{\phi_2}-\vec{\phi_2}\cdot\partial_\mu\vec{\phi_1}\big)^2\nonumber\\ 
&+\frac{W^{(2)}(\rho,\tau)}{4}\Big[\big(\vec{\phi_1}\cdot\partial_\mu\vec{\phi_2}+\vec{\phi_2}\cdot\partial_\mu\vec{\phi_1}\big)^2 \nonumber\\
&+\big(\vec{\phi_1}\cdot\partial_\mu\vec{\phi_1} -\vec{\phi_2}\cdot\partial_\mu\vec{\phi_2}\big)^2\Big]
+\frac{W^{(3)}(\rho,\tau)}{4} S_{a b}\partial_\mu \vec{\phi_a} \cdot \partial_\mu \vec{\phi_b} \nonumber\\
&+\frac{W^{(4)}(\rho,\tau)}{4} S_{a c} (\vec{\phi}_a\cdot\partial_{\mu}\vec{\phi}_b)(\vec{\phi}_b \partial_{\mu} \vec{\phi}_c)\nonumber\\
&+\frac{W^{(5)}(\rho,\tau)}{4} S_{a c} (\vec{\phi}_b\cdot\partial_{\mu}\vec{\phi}_a)(\vec{\phi}_b \partial_{\mu} \vec{\phi}_c) \nonumber\\
&+\frac{W^{(6)}(\rho,\tau)}{4} S_{a b} (\vec{\phi}_b\cdot\partial_{\mu}\vec{\phi}_a) S_{c d} (\vec{\phi}_c\cdot\partial_{\mu}\vec{\phi}_d)
\biggr\}.
\end{align}
In addition to the potential $U_k$, this ansatz involves eight functions, the RG flow equations of which are gigantic. To simplify the problem, we have retained below only the $Z_k$ and $Y_k$ functions which constitutes the minimal set of functions sufficient to fully describe at $O(\partial^2)$ the Gaussian and $O(2N)$ fixed points. However, we include their full dependence on $\rho$ and $\tau$. We will call this approximation $DE2^{ZY}$.

\section{Results}
\label{sec_Results}

In this section, we present the results obtained for various quantities in the $O(N)\times O(2)$ universality class. To begin with, we will briefly present how we estimate central values for the various quantities computed with the LPA and $DE2^{ZY}$ approximations as well as how we find the corresponding error bars. We will then study the $N_c^+(d)$, $N_c^-(d)$ and $N_c^{H}(d)$ curves defined in the introduction and compare them with previous estimates from the literature. We will then calculate the critical exponents in the three-dimensional case for those values of $N$ for which we find a fixed point characterizing a second-order phase transition. Finally, we analyze the possibility of a second order phase transition disconnected from the $N>N_c^+(d)$ region around $N=2$ or $N=3$ in $d=3$.

\subsection{Central values and error bars}
\label{sec_error}

All the quantities we calculate - critical exponents and critical lines - are functions of the $\alpha$ parameter defined in Eq.~(\ref{eq:RegProf}). We assume that the optimal values of these quantities are obtained using the PMS. Strictly speaking, the criterion of convergence of the DE derived in Ref.~\cite{DePolsi2020} cannot be applied here since we do not use the complete $\mathcal{O}(\partial^2)$ ansatz but only the $DE2^{ZY}$ ansatz. We nevertheless assume that all the quantities calculated with this ansatz are more accurate than in LPA, i.e. that they are closer to the exact results. We therefore take as the central values of the calculated quantities those obtained at $DE2^{ZY}$ and as the error bars half the difference between the values at LPA and $DE2^{ZY}$. We show below that this estimate of the error bars seems to overestimate them, but in the absence of more robust criteria we retain them.

Considered as functions of $\alpha$, some quantities are convex at LPA and concave at $DE2^{ZY}$ or vice versa. In this case, the PMS values could be assumed to be lower and upper bounds and the
error bars could be reduced. We choose to not use this assumption to remain as conservative as possible.

\subsection{The $N_c^+(d)$ curve}
\label{secncp}

\begin{figure}[t]
    \centering
    \includegraphics[width=8.4cm]{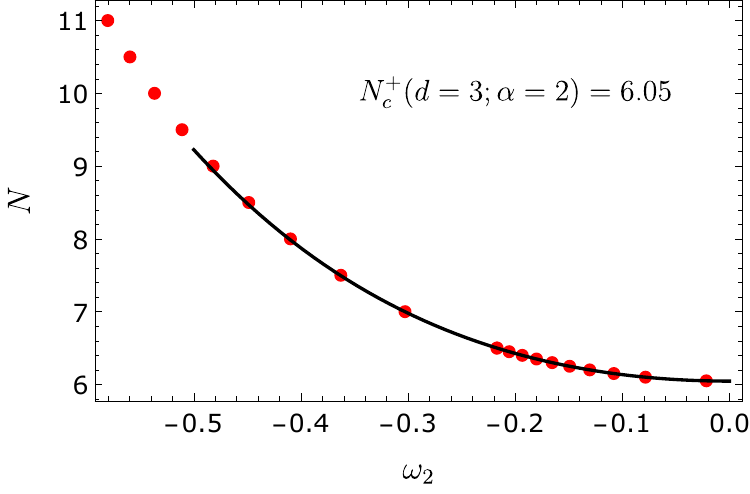}
    \caption{Dependence of $\omega_2$ with $N$ near $N_c^+$ for $d=3$ and $\alpha=2$. The data obtained from the $DE2^{ZY}$ approximation are represented by red dots. 
    The curve $N(\omega_2)$ has an horizontal tangent at $N_c^+$. The precise value of $N_c^+$ is obtained by fitting $N(\omega_2)$ by a quartic curve shown in black.}
    \label{fig:w2vsN}
\end{figure}

We numerically determine the value of $N_c^+$ for a given dimension $d$ by lowering the values of $N$ until the $C_+$ fixed point disappears by colliding with $C_-$. This occurs when the velocity of the flow along the line joining $C_+$ and $C_-$ vanishes, that is, when the first irrelevant eigenvalue $\omega_2$ at $C_+$ vanishes. We therefore look for $C_+$ by the Newton-Raphson method at large enough $N$ and $d$ using a simple $\phi^4$ truncation of the LPA fixed point equation as a first try. Then, for this $d$, we decrease $N$ and plot $\omega_2(N)$, see Fig.~\ref{fig:w2vsN}. Since this curve is steep in $N$ in the vicinity of $N_c^+$, its determination is easy and accurate. We then lower $d$ and repeat the procedure.

\begin{figure}[t]
    \centering
    \includegraphics[width=8.4cm]{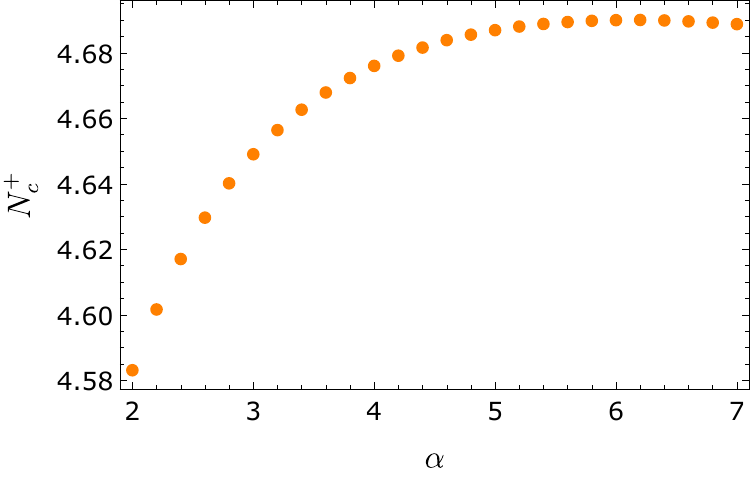}
    \includegraphics[width=8.4cm]{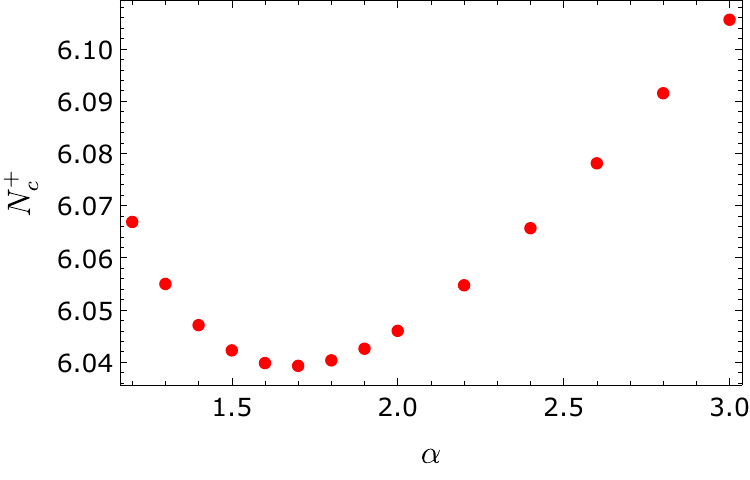}
    \caption{Dependence of $N_c^+$ on $\alpha$ for $d=3$. Upper curve: results in LPA. Bottom curve: results in $DE2^{ZY}$ approximation. As can be seen, the dependence on $\alpha$ of both curves is very small. In particular it is much smaller than the typical difference between $DE2^{ZY}$ and LPA.}
    \label{fig:ncvsalpha}
\end{figure}

\begin{figure}[t]
    \centering
    \includegraphics[width=8.4cm]{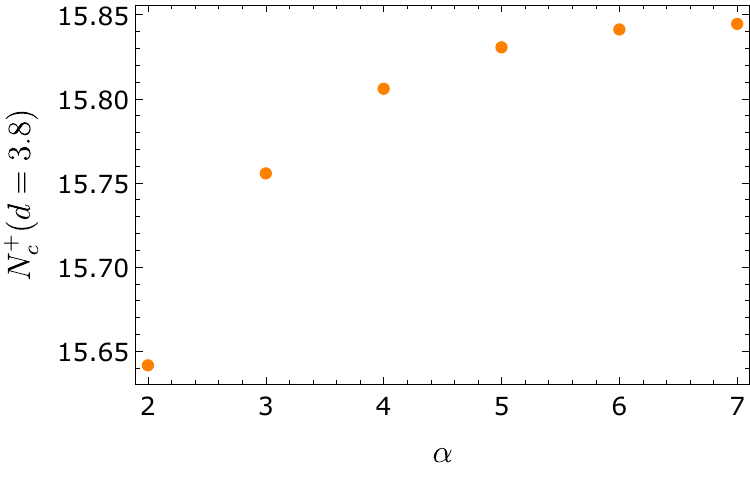}
    \includegraphics[width=8.4cm]{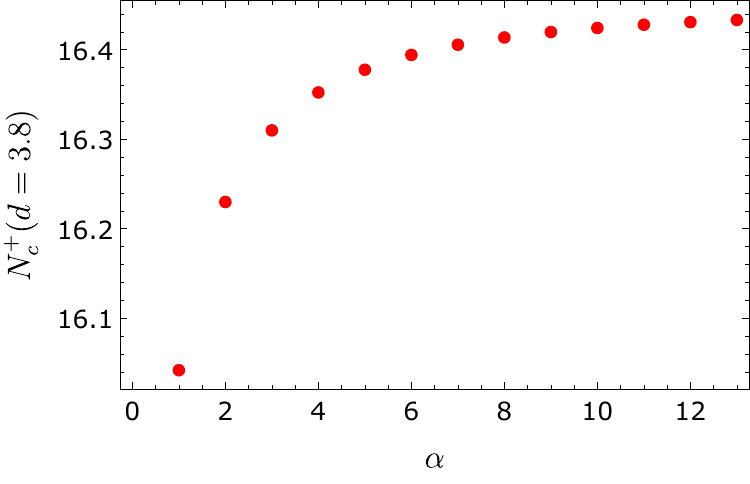}
    \caption{Dependence of $N_c^+$ on $\alpha$ for $d=3.8$. Upper curve: results in LPA. Bottom curve: results in $DE2^{ZY}$ approximation. As can be seen, the dependence on $\alpha$ of both curves is very small. In particular it is much smaller than the typical difference between $DE2^{ZY}$ and LPA.}
    \label{fig:ncvsalphadlarge}
\end{figure}

\begin{figure}[t]
    \centering
    \includegraphics[width=8.4cm]{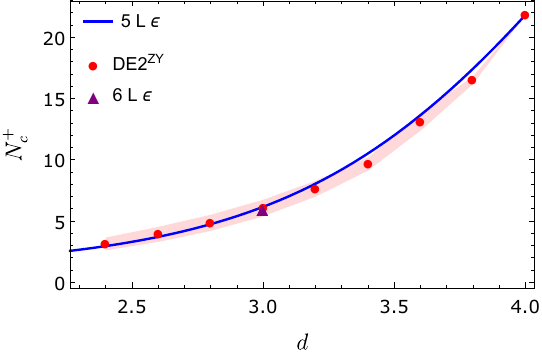}
    \caption{$N_c^+(d)$ obtained in $DE2^{ZY}$ approximation (red dots). The pink zone corresponds to our estimate of error bars
 (central values are given by PMS values of $DE2^{ZY}$ approximation and error bars are given by half difference with LPA result). 
 The blue curve is the $\epsilon$-expansion result at five loops \cite{calabrese2004five}. The triangle is the value of $N_c^+(d=3)$ obtained in $d=3$ at six loops \cite{Kompaniets:2019xez}. The part above the curve corresponds to a second order phase transition and the part below to a first order phase transition.}
    \label{fig:ncvsdDE}
\end{figure} 

Once the value of $N_c^+$ has been determined for a value of $\alpha$ we vary this parameter in order to determine the position of the PMS. In Fig.~\ref{fig:ncvsalpha}, the curves $N_c^+(\alpha)$ are plotted in the vicinity of their extremum both at LPA and at $DE2^{ZY}$. The two curves have opposite concavity suggesting that LPA and $DE2^{ZY}$ give lower and upper bounds estimates of the exact value of $N_c^+$. This is the behavior observed for all studied values of $d<3.5$. For higher values of $d$ a different behavior is observed as shown in Fig.~\ref{fig:ncvsalphadlarge}. For $d>3.5$ the two curves are concave and the optimal value of $\alpha$ increases to very large values.

In Fig.~\ref{fig:ncvsdDE}, we compare our results of the curve $N_c^+(d)$ with the curve obtained in \cite{calabrese2004five} and $N_c^+(3)$ obtained in Ref.~\cite{Kompaniets:2019xez}. As can be seen, the $N_c^+(d)$ curve tends to the exact value at the upper critical dimension $d=4$ but does so below the exact $\epsilon-$expansion curve for $d=4-\epsilon$. This happens in both considered approximations, suggesting that in this limit our error bar estimates can be slightly underestimated.  Moreover, the region $d>3.5$ where our error bars for $N_c^+(d)$ seem to be underestimated coincide with those where an important change in the $\alpha$ dependence is observed with respect to $d<3.5$. Both phenomena could be related and this may require a specific study in the future. In any case, it should be noted that the discrepancy is very small (our central values never differ from the curve obtained by the $\epsilon$-expansion by more than 10\%).

In spite of that, it is worth noting that the curve to order $DE2^{ZY}$ in Fig.~\ref{fig:ncvsdDE} is very similar to others obtained in the literature \cite{Delamotte:2010ba,Reehorst:2024vyq,Kompaniets:2019xez}. 
Furthermore, again comparing with other results in the literature, the $DE2^{ZY}$ approximation shows significantly better results than the LPA.
In fact, at $d=3$ one obtains $N_c^+(d=3)=6.0(7)$ which is very similar to results obtained by other approximation schemes. An important point is that we have been able to reach much lower dimensions than $d=3$ without observing any indication of an ``S'' behavior. 
Moreover, not only is the curve not S-shaped, but it is also does not show a change a concavity, making the possibility of a return point even more remote.
In this respect, albeit with a lower level of rigor, our work reaches a regime not yet explored by the Conformal Bootstrap \cite{Reehorst:2024vyq}. 

To sum up, this analysis confirms and improves the one previously obtained by the LPA or LPA' approximations in the FRG framework. The present work does not observe any indication of ``S'' behavior of the $N_c^+(d)$ curve, and allows to have a rough estimate of error bars. The value $N_c^+(d=3)$ is compatible with 6, which is much larger than the physically interesting values $N=2$ and $3$ suggesting that the transition is first order for these values of $N$.

\subsection{The $N_c^-(d)$ and $N_c^{H}(d)$ curves}

In this paper, in addition to analyzing the $N_c^{+}(d)$ curve, we also analyze the $N_c^{-}(d)$ and $N_c^{H}(d)$ curves that have been much less studied in general and have never been studied with the FRG.

\begin{figure}[ht]
    \centering
    \includegraphics[width=8.4cm]{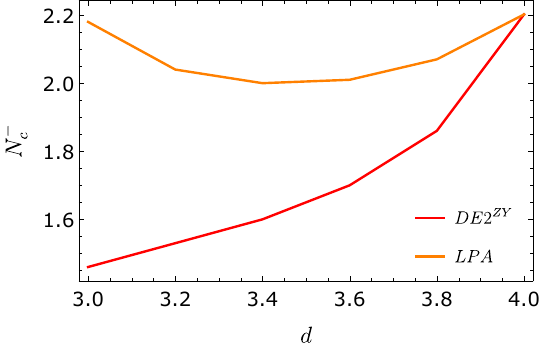}
        \caption{$N_c^-(d)$ obtained in LPA (orange curve) and $DE2^{ZY}$ (red curve). All the values are calculated for $\alpha=2$. The part below the curve corresponds to a region of second order phase transition and the region above the curve to first order phase transitions.} 
    \label{fig:ncmvsdDE}
\end{figure}

\begin{figure}[ht]
    \centering
    \includegraphics[width=8.4cm]{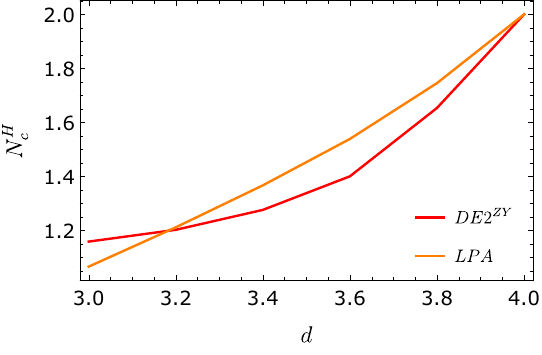}
    \caption{$N_c^{H}(d)$ obtained in LPA (orange curve) and $DE2^{ZY}$ (red curve). All the values are calculated for $\alpha=2$. The part below the curve corresponds to a second order phase transition and the part above to a first order phase transition.}
    \label{fig:ncHvsdDE}
\end{figure}

We show these curves in Figs.~\ref{fig:ncmvsdDE},\ref{fig:ncHvsdDE} and for $\alpha=2$ which is the unique  value we have studied and that we have chosen because it is close to the PMS value for other quantities, see Secs.~\ref{secncp} and \ref{seccritexp} (see below for the study of the $\alpha$-dependence in $d=3$).

\begin{figure}[ht]
    \centering
    \includegraphics[width=8.4cm]{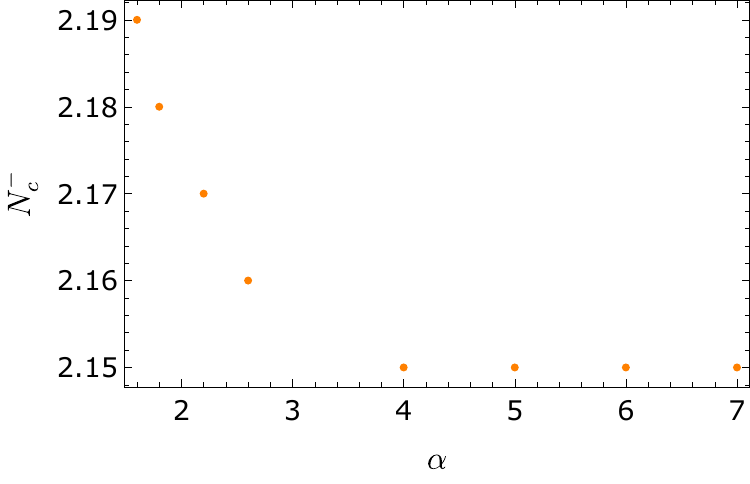}
    \includegraphics[width=8.4cm]{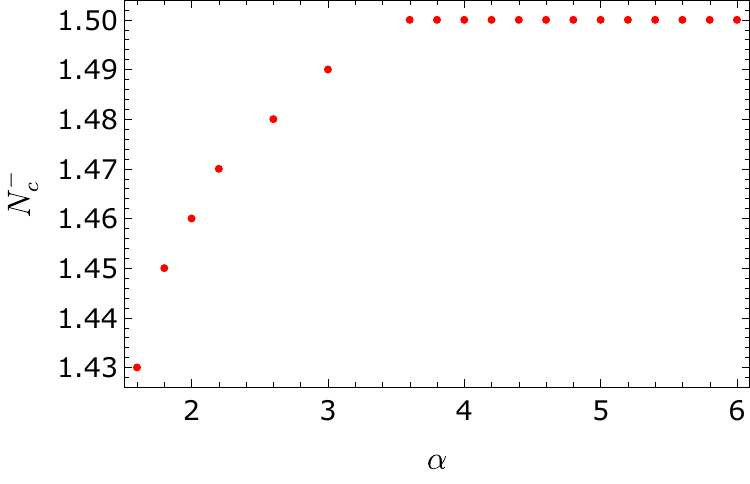}
        \caption{$N_c^-(d=3)$ as a function of $\alpha$ obtained in LPA (top) and $DE2^{ZY}$ (bottom). All the values are calculated for $\alpha=2$. As can be seen, the dependence on $\alpha$ of both curves is very small. In particular it is much smaller than the typical difference between $DE2^{ZY}$ and LPA.}
    \label{fig:ncmvsalphaDE}
\end{figure}

\begin{figure}[ht]
    \centering
    \includegraphics[width=8.4cm]{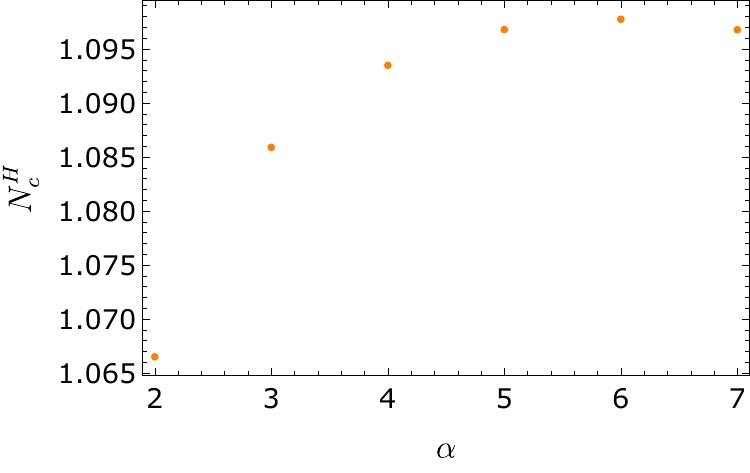}
    \includegraphics[width=8.4cm]{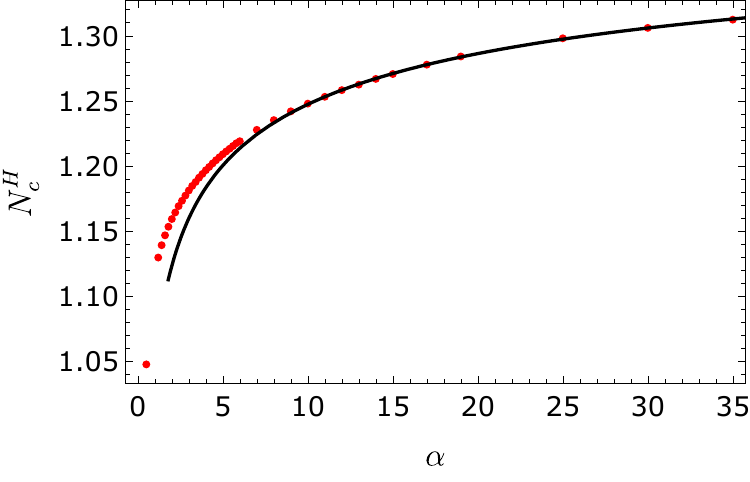}
        \caption{$N_c^{H}(d=3)$ as a function of $\alpha$ obtained in LPA (top) and $DE2^{ZY}$ (bottom). In the LPA case one observes a PMS for $\alpha\simeq 6$. In the case of the $DE2^{ZY}$ approximation, the stability seems to take place for asymptotically large values of $\alpha$. We show with a continuous black line the fit $N_c^{H}(d=3)\simeq 1.47666 -0.425975\times x^{-0.268779}$ that is obtained for large values of $\alpha$. As can be seen, the dependence on $\alpha$ of both curves is very small. In particular it is much smaller than the typical difference between $DE2^{ZY}$ and LPA.}
    \label{fig:ncHvsalphaDE}
\end{figure}

We have studied the dependence of $N_c^-$ and $N_c^{H}$ on $\alpha$ in $d=3$, see Figs.~\ref{fig:ncmvsalphaDE} and \ref{fig:ncHvsalphaDE}. We observe that although this dependence is not small, a PMS is found at asymptotically large values of $\alpha$, the variations becoming very small for $\alpha$ typically larger than 4 for both the LPA and $DE2^{ZY}$. We also notice that the variations of $N_c^-(d=3)$ and $N_c^{H}(d=3)$ with $\alpha$ are significantly smaller than the difference between the LPA and $DE2^{ZY}$ values.
Using the procedure explained in Sec.~\ref{sec_error}, we obtain $N_c^-(d=3) = 1.5(3)$. Notice that this result is not compatible with $N_c^-(d =3) = 1.970(3)$ obtained in \cite{Kompaniets:2019xez}
in the $\epsilon = 4-d$
expansion at six loops. This implies that the error bars, whether the ones we give or the ones in \cite{Kompaniets:2019xez}, may be underestimated. In contrast, we obtain $N_c^H(d = 3) = 1.48(9)$, which is compatible with the result $N_c^H(d = 3) = 1.462(13)$ from \cite{Kompaniets:2019xez}.

In general terms, the results obtained for $N_c^-(d=3)$
and $N_c^H(d=3)$ with these levels of approximation coincides qualitatively with the results obtained in perturbation theory \cite{Kompaniets:2019xez}. A more detailed analysis of these critical values of $N$ is probably desirable but goes beyond the exploratory studies addressed in the present section.

\subsection{Critical exponents in $d=3$}
\label{seccritexp}

\begin{table}[ht]
\caption{\label{expN7final} Results 
at LPA and $DE2^{ZY}$ for
$N=7$ in $d=3$. 
For comparison, results from LPA'  \cite{Delamotte:2015zgf}, MC \cite{Sorokin:2022zwh}, 
6-loop, $d=3$ \cite{Calabrese:2003ib}, 5-loop expansion in MS scheme \cite{Calabrese:2004nt}, 5-loop $\epsilon-$expansion
\cite{calabrese2004five},
6-loop $\epsilon-$expansion \cite{Kompaniets:2019xez}
are also given for comparison and large$-N$ expansion \cite{Pelissetto:2001fi,Gracey:2002ze}.}
\begin{ruledtabular}
\begin{tabular}{lllll}
                 &   $\nu$         &  $\eta$      &  $\omega_1$ &  $\omega_2$\\
\hline
LPA\,              &   0.80        &  0          &  0.85      & 0.56   \\
$DE2^{ZY}$\,       &   0.71(5)     &  0.047(24)  &  0.89(2)   & 0.30(12) \\
      \hline
LPA'      & 0.74 &  0.039 & & 0.48\\
MC      & 0.739(7) &  0.030(17)  &    \\
6-loop, $d=3$     & 0.68(2)		&  & 0.83(2) & 0.23(5) \\
5-loop, $d=3$ $\overline{MS}$    & 0.68(4)		& 0.047(15) & 0.8(2) & 0.5(2)	 \\
5-loop, $\epsilon-$expansion & 0.71(4)	 &  & 0.84(3) & 0.33(10)	\\
6-loop, $\epsilon-$expansion & 0.713(8)	 & 0.045(3) & 0.812(13) & 0.34(2)	\\
large$-N$             &0.70(4)      & 0.053(2)      & 0.77(12) & 0.54(23)
\end{tabular}
\end{ruledtabular}
\end{table}
	
\begin{table}[]
\caption{\label{expN8final} Results 
at LPA and $DE2^{ZY}$ for
$N=8$ in $d=3$. 
For comparison, results from MC \cite{Sorokin:2022zwh}, 
and 6-loop, $d=3$ perturbative \cite{Calabrese:2003ib}, 5-loop expansion in MS scheme \cite{Calabrese:2004nt}, 5-loop $\epsilon-$expansion
\cite{calabrese2004five},
6-loop $\epsilon-$expansion \cite{Kompaniets:2019xez} and large$-N$ expansion \cite{Pelissetto:2001fi,Gracey:2002ze}
are also given for comparison.}
\begin{ruledtabular}
\begin{tabular}{lllll}
                 &   $\nu$         &  $\eta$      &  $\omega_1$ &  $\omega_2$\\
\hline
LPA\,              &   0.83        &  0          &  0.87   & 0.63   \\
$DE2^{ZY}$\,       &   0.74(5)     &  0.043(22)  &  0.89(1)& 0.41(11)   \\
      \hline
MC      & 0.771(8)  &  0.034(20)  &    \\
6-loop $d=3$     & 	0.71(1)	&  & 0.83(2) & 0.36(4) \\
6-loop $d=3$ $\overline{MS}$  & 	0.70(5)	& 0.042(10) & 0.8(2) & 0.5(2) \\
5-loop, $\epsilon-$expansion & 0.75(4)	 &  & 0.84(3) & 0.45(8)	\\
6-loop, $\epsilon-$expansion & 0.745(11)	 & 0.042(2) & 0.81(4) & 0.447(15)	\\
large$-N$               &0.74(3)      & 0.047(2)     & 0.80(10) & 0.59(20)
\end{tabular}
\end{ruledtabular}
\end{table}

\begin{table}[]
\caption{\label{expN10final} Results 
at LPA and $DE2^{ZY}$ for
$N=10$ in $d=3$. 
For comparison, results from CB \cite{Henriksson:2020fqi}, 6-loop, $d=3$ \cite{Calabrese:2003ib}, 5-loop expansion in MS scheme \cite{Calabrese:2004nt} and large$-N$ expansion \cite{Pelissetto:2001fi,Gracey:2002ze}
are also given for comparison.}
\begin{ruledtabular}
\begin{tabular}{lllll}
                 &   $\nu$         &  $\eta$      &  $\omega_1$ &  $\omega_2$\\
\hline
LPA\,              &   0.87       &  0          &  0.892   & 0.73   \\
$DE2^{ZY}$\,       &   0.79(4)    &  0.039(20)  &  0.897(3) & 0.54(9)  \\
      \hline
CB      &  0.71 (8)  &  0.050 (24) &     \\
6-loop $d=3$ $\overline{MS}$  & 	0.74(5)	& 0.036(8) & 0.9(2) & 0.5(2) \\
large$-N$               &0.80(2)       & 0.038(1)    & 0.84(8) & 0.68(16)
\end{tabular}
\end{ruledtabular}
\end{table}
		
\begin{table}[]
\caption{\label{expN16final} Results 
at LPA and $DE2^{ZY}$ for
$N=16$ in $d=3$. 
For comparison, results from LPA'-field expanded \cite{Delamotte:2003dw},
6-loop, $d=3$ \cite{Calabrese:2003ib}, 5-loop expansion in MS scheme \cite{Calabrese:2004nt}, 5-loop $\epsilon-$expansion
\cite{calabrese2004five}, 6-loop $\epsilon-$expansion
\cite{Kompaniets:2019xez} and large$-N$ expansion \cite{Pelissetto:2001fi,Gracey:2002ze}
are also given for comparison.}
\begin{ruledtabular}
\begin{tabular}{lllll}
                 &   $\nu$         &  $\eta$      &  $\omega_1$ &  $\omega_2$\\
\hline
LPA\,              &   0.92      &  0          &  0.934   & 0.84   \\
$DE2^{ZY}$\,       &   0.86(3)   &  0.028(14)  &  0.922(3) & 0.71(7)  \\
      \hline
LPA'-FE      & 0.898 & 0.0252 &\\
6-loop $d=3$     & 	0.863(4)	&  &  0.876(4) & 0.714(9)\\
6-loop $d=3$ $\overline{MS}$    & 	0.82(4)	& 0.025(4) &  0.9(2) & 0.74(12)\\
5-loop, $\epsilon-$expansion     & 	0.89(4) &   & 0.86(1) & 0.77(2)	\\
6-loop, $\epsilon-$expansion     & 	0.850(16) & 0.0261(7) & 0.860(3) & 0.771(6)	\\
large$-N$               &0.885(7)       & 0.0245(4)  &  0.90(5) & 0.80(10)
\end{tabular}
\end{ruledtabular}
\end{table}

\begin{table}[]
\caption{\label{expN32final} Results 
at LPA and $DE2^{ZY}$ for
$N=32$ in $d=3$. 
For comparison, results from LPA'-field expanded \cite{Delamotte:2003dw}, 6-loop, $d=3$ \cite{Calabrese:2003ib}, 5-loop $\epsilon-$expansion
\cite{calabrese2004five}, 6-loop $\epsilon-$expansion
\cite{Kompaniets:2019xez} and large$-N$ expansion \cite{Pelissetto:2001fi,Gracey:2002ze}
are also given for comparison.}
\begin{ruledtabular}
\begin{tabular}{lllll}
                 &   $\nu$         &  $\eta$      &  $\omega_1$ &  $\omega_2$\\
\hline
LPA\,              &   0.96      &  0          &  0.968    & 0.93   \\
$DE2^{ZY}$\,       &   0.93(2)   &  0.016(8)   &  0.958(5) & 0.86(4)   \\
      \hline
LPA'-FE       & 0.95 & 0.0134 &\\
6-loop, $d=3$     & 	0.936(1)	&  & 0.933(2) & 0.868(2)\\
5-loop, $\epsilon-$expansion     & 0.94(2)	 &  & 0.91(2)	& 0.90(1)\\
6-loop, $\epsilon-$expansion     & 0.940(17)	 & 0.014(3) & 0.921(10)	& 0.909(7)\\
large$-N$                &0.946(2)     & 0.0125(1)   & 0.95(3)  & 0.90(5)
\end{tabular}
\end{ruledtabular}
\end{table}

\begin{table}[]
\caption{\label{expN6final} Results 
at LPA for $N=6$ in $d=3$. 
For comparison, results from LPA' \cite{Delamotte:2015zgf}, LPA'-field expanded \cite{Delamotte:2003dw}, CB \cite{Henriksson:2020fqi}, MC \cite{loison2000critical,Sorokin:2022zwh}, 5-loop expansion in MS scheme \cite{Calabrese:2004nt} and 
6-loop $\epsilon-$expansion \cite{Kompaniets:2019xez}
are also given for comparison.}
\begin{ruledtabular}
\begin{tabular}{lllll}
                 &   $\nu$         &  $\eta$      &  $\omega_1$ &  $\omega_2$\\
\hline
LPA\,              &   0.76 &  0          &  0.84   & 0.45   \\
      \hline
LPA'                    &   0.70 & 0.042 &  & 0.33 \\
LPA'-FE                  &   0.707     & 0.053     &  &       \\
CB      &  0.79(7) &	0.033(8)     &     \\
MC Loison                      &  0.700(11) &  0.025(20)    \\
MC Sorokin                     &  0.686(7) &  0.032(17)    \\
5-loops, $d=3$ $\overline{MS}$     & 0.66(4)		& 0.052(14) & \\
6-loops, $\epsilon-$expansion & 0.65(2)	 & 0,047(3) & 0.83(2)
\end{tabular}
\end{ruledtabular}
\end{table}


As it results from the previous analysis, our study agrees with several other previous works on the existence of a second order phase transition in $d=3$ for $N \gtrsim 6$. It is therefore worthwhile to analyze the values obtained for the
critical exponents for such values of $N$ as a means of testing the
quality of the approximations employed, and to compare them with values
previously reported in the literature.  We do this
in Tables~\ref{expN7final} to \ref{expN6final}, for the exponents $\eta$, $\nu$, $\omega_1$ and $\omega_2$ and for different
values of N. For $N=6$, no FP is found for the $DE2^{ZY}$ approximation and only the LPA values are given, see Table \ref{expN6final}.
In Tables~\ref{expN7final} to \ref{expN32final}, we compare our results with those
obtained in the large-$N$ expansion that are given by  \cite{Pelissetto:2001fi,Gracey:2002ze}:
\begin{align}
 \nu&=1-\frac{16}{\pi^2 N}-\frac{8(21\pi^2 - 80)}{3\pi^4 N^2} +\mathcal{O}(1/N^3)\nonumber\\
 \eta&=\frac{4}{\pi^2 N}-\frac{64}{3\pi^4 N^2}+\mathcal{O}(1/N^3) \nonumber\\
 \omega_1&=1-\frac{16}{\pi^2 N} +\mathcal{O}(1/N^2)\nonumber\\
 \omega_2&=1-\frac{32}{\pi^2 N} +\mathcal{O}(1/N^2).
\end{align}
These formulas are employed in the following way: we use as central value the expression for the exponents at the largest available order ($1/N^2$ for $\nu$ and $\eta$ and $1/N$ for $\omega_1$ and $\omega_2$) and we employ half the difference between that order and the previous one as an estimate of error bar in the corresponding exponent.

\begin{figure}[ht]
    \centering
    \includegraphics[width=8.4cm]{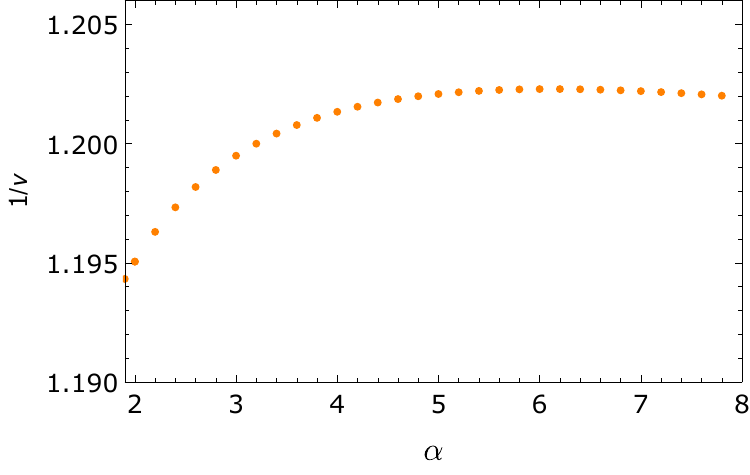}
    \includegraphics[width=8.4cm]{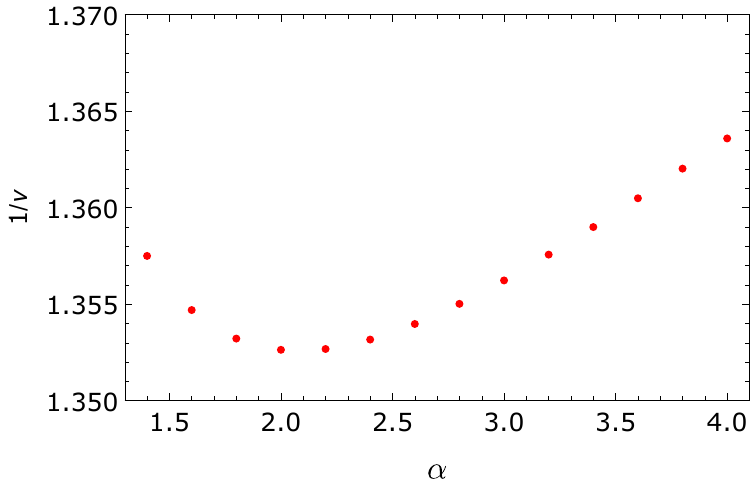}
    \caption{Dependence of the exponent $\nu$ on $\alpha$ for $d=3$ and $N=8$. Upper curve: results in LPA. Bottom curve: results in $DE2^{ZY}$ approximation. As can be seen, the dependence on $\alpha$ of both curves is very small. In particular it is much smaller than the typical difference between $DE2^{ZY}$ and LPA.}
    \label{fig:nuvsalpha}
\end{figure} 

\begin{figure}[ht]
    \centering
    \includegraphics[width=8.4cm]{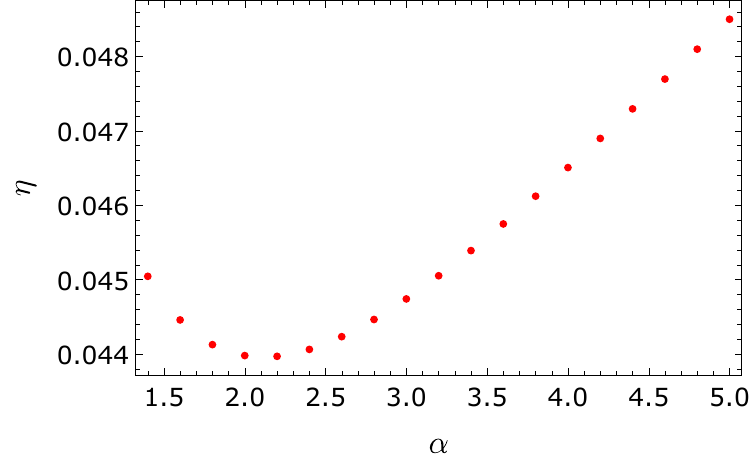}
    \caption{Dependence of the exponent $\eta$ on $\alpha$ for $d=3$ and $N=8$. Results in $DE2^{ZY}$ approximation.}
    \label{fig:etavsalpha}
\end{figure}

\begin{figure}[h]
    \centering
    \includegraphics[width=8.4cm]{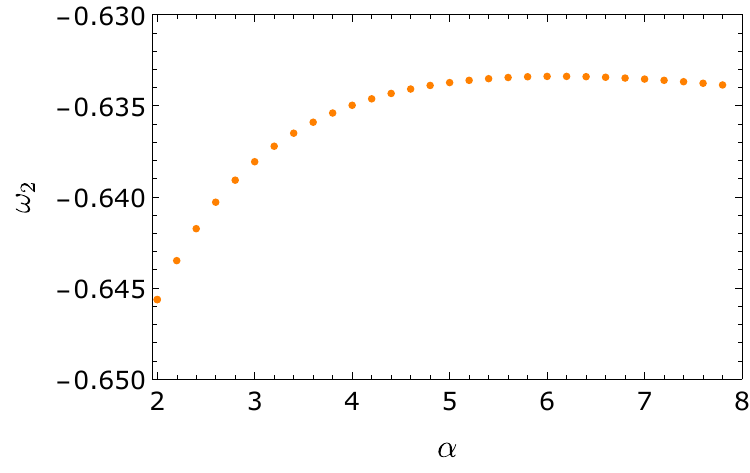}
    \includegraphics[width=8.4cm]{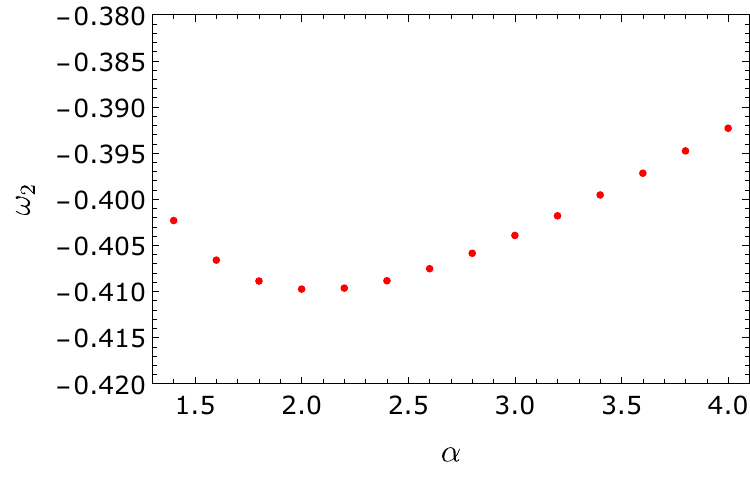}
    \caption{Dependence of the exponent $\omega_2$ on $\alpha$ for $d=3$ and $N=8$. Upper curve: results in LPA. Bottom curve: results in $DE2^{ZY}$ approximation. As can be seen, the dependence on $\alpha$ of both curves is very small. In particular it is much smaller than the typical difference between $DE2^{ZY}$ and LPA.}
    \label{fig:omega2vsalpha}
\end{figure}

\begin{figure}[ht]
    \centering
    \includegraphics[width=8.4cm]{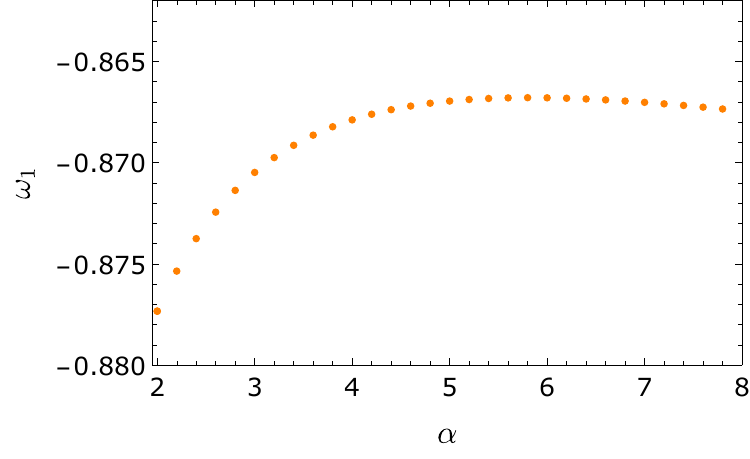}
    \includegraphics[width=8.4cm]{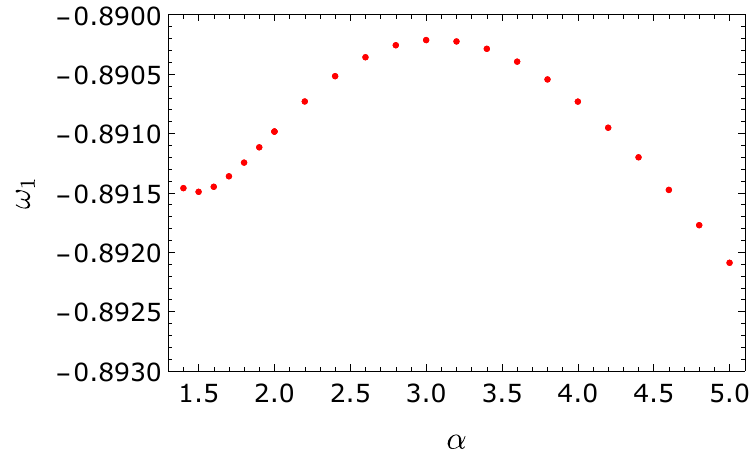}
    \caption{Dependence of the exponent $\omega_1$ on $\alpha$ for $d=3$ and $N=8$. Upper curve: results in LPA. Bottom curve: results in $DE2^{ZY}$ approximation. As can be seen, the dependence on $\alpha$ of both curves is very small. However, it is not much smaller than the typical difference between $DE2^{ZY}$ and LPA. The extremum selected at $DE2^{ZY}$ approximation is the one that minimizes the difference with the LPA, that is, corresponds to $\alpha=3.0$.}
    \label{fig:omega1vsalpha}
\end{figure}

As can be seen for $N=8$ in Figs.~\ref{fig:nuvsalpha} to \ref{fig:omega2vsalpha}, the concavities of the $\nu(\alpha)$, $\eta(\alpha)$ and $\omega_2(\alpha)$ functions change between LPA and $DE2^{ZY}$ which suggests that the PMS values of these quantities are lower and upper bounds. This turns out to be true and to be general for all the values of $N$ we have studied. This is consistent with
the results in Tables \ref{expN7final} to \ref{expN32final}: both approximations are alternating bounds
for these exponents. Only results for $N=8$ are shown in Figs. 
\ref{fig:nuvsalpha}, \ref{fig:etavsalpha} and \ref{fig:omega2vsalpha}  but we have checked that similar behaviors are observed for all values of $N$ studied. On the other hand, the concavity of
$\omega_1(\alpha)$ does not change between LPA and $DE2^{ZY}$, see Fig.~\ref{fig:omega1vsalpha}. Moreover, for this exponent in some cases we obtain more than one PMS point. In such a case we adopted the criterion of choosing the one that makes the difference between LPA and $DE2^{ZY}$ smaller. Again, such a figure is made for $N=8$ but a similar behavior is observed for all values of $N$ studied.

The value $N=1$, deserves a specific mention. First of all it is very important to point out that this case {\it does not correspond} to the Ising STA model.
In fact, the $N=1$ belongs to the $O(2)$ universality class previously studied in the FRG literature. We verified  that our results coincide with those of the $O(2)$ case.

In general, the agreement of the $DE2^{ZY}$ approximation with other results in the literature for critical exponents tends to be much better than that of the LPA. Moreover, for several quantities (and most particularly for $\eta$) the agreement is much better than suggested by the error bars. This indicates that we may be over-estimating them. At the same time, this substantial improvement between the LPA and the $DE2^{ZY}$ approximation in the exponents for $N>7$ is a strong indication that the two-derivative terms that were included give the dominant effect at that order and is an indicator that the approximation  implemented here gives accurate results.

\subsection{The possibility of a second order island}
\label{island}
In this section, we analyze again the possibility of a second order phase transition for $N=2$ and $N=3$. As previously mentioned, our $N_c^+(d)$ curve rules out, at least within the range of validity of our approximations, the possibility of an ``S'' behavior even with a return point significantly lower than $d=3$. However, it has been suggested (for a recent reference see, e.g. \cite{Reehorst:2024vyq}) that the existence of a second-order phase transition could be due to the existence of a second-order ``island'' in the first-order zone. Obviously, the analysis of the $N_c^+(d)$ curve does not allow us to discuss this point and we propose in this section to address it by a complementary analysis.

It should be noted that to completely rule out the existence of a fixed point in the framework of the renormalization group is a difficult task. This is because the space of possible theories (even in an approximate framework) in which evolves the renormalization group flow is a space parameterized by an infinite set of coupling constants. It should be noted, however, that the FRG, both to order LPA and to order $\mathcal{O}(\partial^2)$ has been employed in the past for the purposes of enumerating virtually {\it all} fixed points present with a given field content and symmetry (see, e.g., \cite{Morris:1994jc,Defenu:2017dec}). This includes not only the search for critical fixed points but even, for example, the listing of the infinite set of multicritical fixed points in Ising's universality class in $d=2$. It has also been employed to classify the set of possible perturbative fixed points near the upper critical dimension \cite{Codello:2020lta}. Complementarily, it has been employed in a successful way to search for {\it non-perturbative} fixed points even in supposedly well-understood models such as the $O(N)$ models \cite{Fleming:2020qqx,Yabunaka:2023jlf}. Despite these successes, claiming that a certain model does not possess a fixed point using these methods is, in general, very difficult.

However, for the purposes of studying the models with $O(N)\times O(2)$ symmetry, much less is required than the complete classification of the fixed points of such a model. Indeed, what we are really interested in for the purposes of the current controversy is the possible existence of a {\it critical} (not a {\it multicritical}) fixed point. That is, we are looking for a fixed point whose determination requires the fine-tuning of a single control parameter. Moreover, for the purpose of explaining experiments and/or numerical simulations, the existence of a critical fixed point with a relatively large basin of attraction is required: a critical fixed point that would be attractive within the critical surface only for a small range of parameters would be irrelevant (in the colloquial sense) for all practical purposes.

The analysis of the possible existence of a critical fixed point for $N=2$ and $N=3$ was already performed previously at order LPA and LPA' \cite{Delamotte:2003dw,Delamotte:2010ba}. The procedure employed is the same as the one performed for standard models such as $O(N)$ ones (see Ref.~\cite{Dupuis:2020fhh}): one takes, for example, the most general renormalizable Hamiltonian compatible with the symmetries, chooses relatively arbitrary values for all coupling constants except one (typically the mass parameter) which is adjusted by bipartition until a fixed point is reached (or not). Such a procedure leads in a simple way in LPA or LPA' to fixed points for $N>N_c^+(d)$ and does not lead to a fixed point for $N=2$ or $N=3$ (even taking different possible values of the quartic constants). In such cases it was observed that the flow leads to a region of slow but non-zero flow  before
undertaking a runaway which is the signal of a first order transition.

\begin{figure}[ht]
    \centering
    \includegraphics[width=8.4cm]{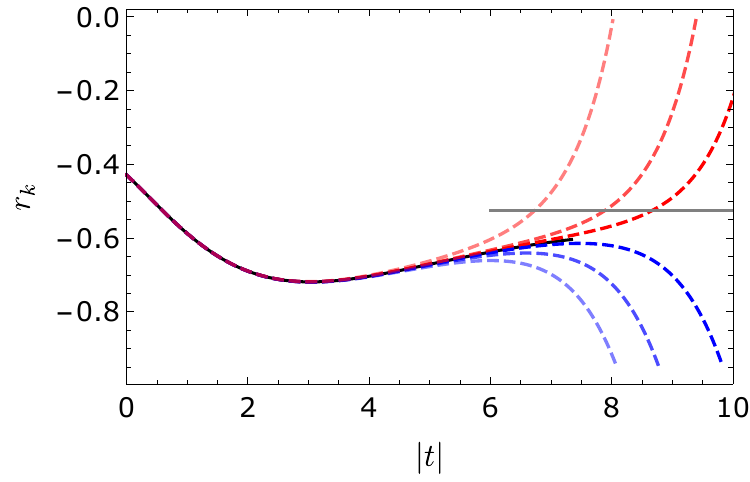}
    \includegraphics[width=8.4cm]{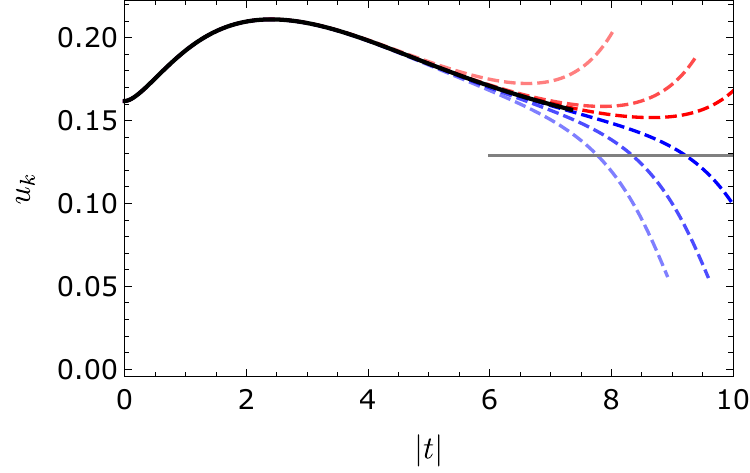}
    \includegraphics[width=8.4cm]{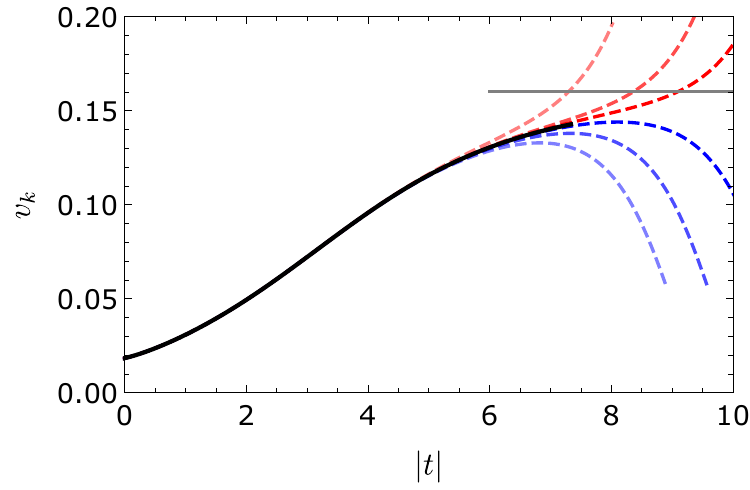}
    \caption{
    Flows of the dimensionless quadratic ($r_k$) and quartic  ($u_k$ and $v_k$) couplings for different initial conditions, see Eq.~(\ref{hamilitonian}) for the definitions of the couplings for $d=3$, $N=7$ and $\alpha=2$.  All initial conditions are close to the critical hypersurface, some being in the low $T$ phase (blue) and the others in the symmetric phase (red). The last flow in the bipartition is shown in black.
    In all cases the grey line indicates the fixed point values of these couplings, that is, the values reached at infinite RG time if the flow were initialized exactly on the critical surface.}
    \label{fig:bipartitionn7}
\end{figure}

 In this work, we have repeated with the $DE2^{ZY}$ approximation the analysis
explained in the previous paragraph. We have verified that the bipartition procedure leads smoothly to the critical fixed point at $d=3$ for $N=7$, $N=8$ and also for $N=1.4$ but does not lead to a fixed point for $N=2$ or $N=3$. We studied the evolution of the full $DE2^{ZY}$ approximation but for representation purposes we studied quadratic and quartic couplings that we defined in the following way:
\begin{align}
		r_k &=\left. \frac{\partial U_k(\rho,\theta)}{\partial \rho}\right\vert_{\rho = 0},
    \hspace{.5cm} u_k = 3\left. \frac{\partial^2 U_k(\rho,\theta)}{\partial \rho^2}\right\vert_{\rho = 0, \theta = \pi/4},\nonumber\\
		v_k &= \frac{3}{2}\Bigg(\left.\frac{\partial^2 U_k(\rho,\theta)}{\partial \rho^2}\right\vert_{\rho = 0, \theta = 0}-\left.\frac{\partial^2 U_k(\rho,\theta)}{\partial \rho^2}\right\vert_{\rho = 0, \theta = \pi/4}\Bigg).
		\label{eqn:frust_var}
\end{align}
The running of these couplings is shown in Fig.~\ref{fig:bipartitionn7} for $N=7$.

\begin{figure}[ht]
    \centering
    \includegraphics[width=8.4cm]{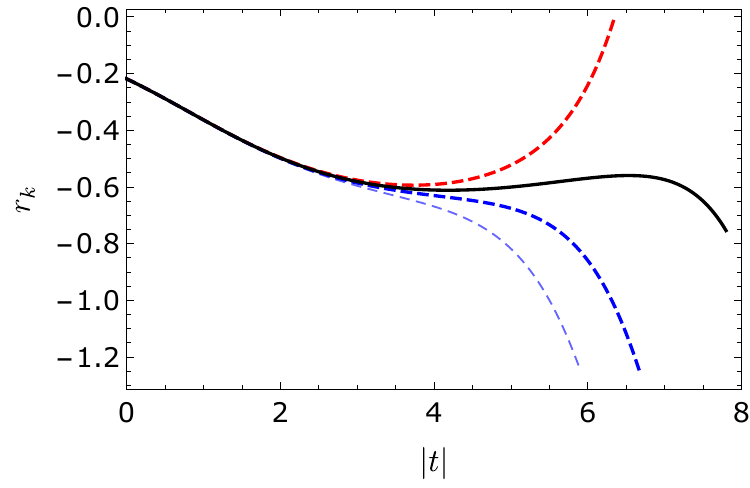}
    \includegraphics[width=8.4cm]{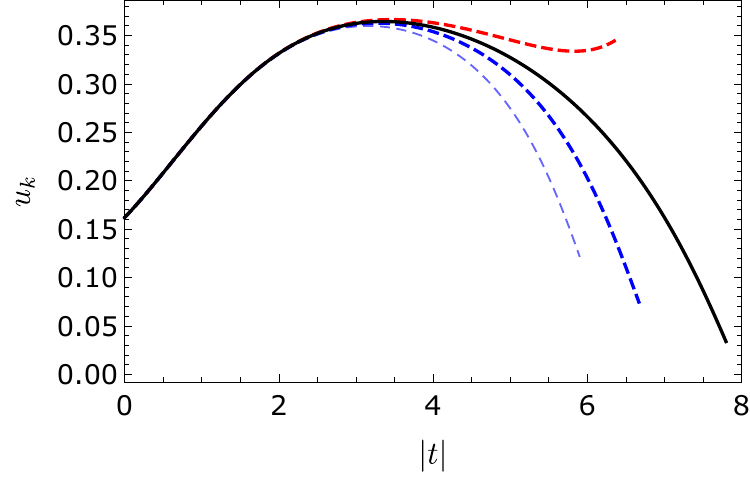}
    \includegraphics[width=8.4cm]{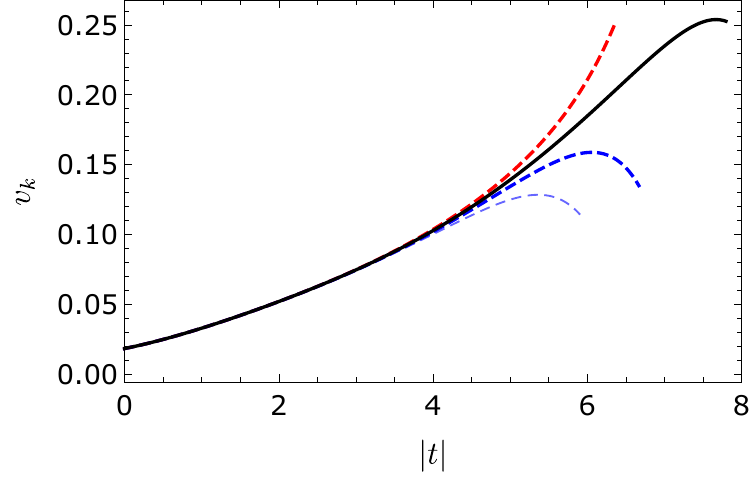}
    \caption{Flows of the dimensionless quadratic ($r_k$) and quartic  ($u_k$ and $v_k$) couplings for different initial conditions, see Eq.~(\ref{hamilitonian}) for the definitions of the couplings for $d=3$, $N=3$ and $\alpha=2$.  All initial conditions are close to the critical hypersurface, some being in the low $T$ phase (blue) and the others in the symmetric phase (red).  The last flow in the bipartition is shown in black.}
    \label{fig:bipartitionn3}
\end{figure}

As usual, the bipartion procedure is not very precise for finding a fixed point
(and an order of magnitude numerically more costly than the FP tracking presented before) but once a reasonable approximation of the fixed point is obtained this way, it can be used as the starting point of a Newton-Raphson algorithm.
The fact that we find a fixed point by this method shows, as expected, 
that the natural initial conditions of the RG flow we are using are
indeed in the basins of attraction of the fixed points for these values of $N$.
On the contrary, for $N=2$ and $N=3$, the same kind of initial conditions never lead to a fixed point.
For example, the flows for the quadratic and quartic couplings for $N=3$ are shown in Fig.~\ref{fig:bipartitionn3}. It is observed that a large number of iterations can be done approaching to a slow-running region but, finally, the anisotropic dimensionless coupling $v_k$ tend to grow and the isotropic coupling $u_k$ tend to approach zero. When $u_k$ becomes negative the bipartition procedure is no longer stable and no fixed point is found.

To ensure that we
are not considering an incorrect region of initial conditions for $u$ and
$v$, we have repeated the procedure by considering the two possible signs of $v$. As previously obtained  in the literature with the LPA', the flow is
always driven to a region where it is slow before undertaking a runaway
which is the signal a first-order transition.

Although the previous analysis does not categorically rule out the existence of a critical fixed point for $d=3$ and $N=2$ or $N=3$, it makes the possibility of an ``island'' explaining the experimental or numerical results rather implausible (at least in the range of validity of our approximation). Indeed, for this to happen, the mere existence of a fixed point would not be sufficient, but would require it to be reached in a relatively natural way by adjusting a single control parameter (such as temperature). The previous analysis shows that if such a fixed point exists, it is extremely elusive, undoubtedly not very ``attractive'' according to the renormalization flow and, therefore, incapable of playing a relevant role.  

\section{Conclusions}\label{secConcl}

In this paper we have analyzed the Ginzburg-Landau model with $O(N)\times O(2)$ symmetry, which is believed to describe the phase transition of some frustrated magnetic systems for $N=2$ and $N=3$. As presented throughout the paper, there has been a controversy about the nature, first or second order, of such a transition for more than two decades. Several papers, both theoretical and experimental, argue that such a transition is second order while a comparable set argues that it is weakly first order.

In this controversy, the studies carried out using the Functional Renormalization Group have played a central role, arguing that there is no critical fixed point for $d=3$ and $N=2$ or $3$ and, therefore, that the transition is (weakly) first order.
These works have been performed at the leading order of the approximation scheme known as Derivative Expansion, or its improved versions. However, since the controversial nature of the phase transition remains unresolved, even with a variety of methodologies, it has become essential to analyze the extent to which higher-order contributions in the Derivative Expansion may or may not alter the previous results.

We analyze for the first time the role of the most relevant functions appearing at the second order of the Derivative Expansion for the $O(N)\times O(2)$ model. While taking into account the full set of possible functions is a very challenging task that cannot be tackled today directly, including the main contributions allows us to improve and study the stability of the results obtained at leading order and to estimate error bars. As for $N\gtrsim 6$ and $d=3$, the critical exponents obtained with different methodologies are fully consistent. In the present work, we use this region of the $(d,N)$ plane to test our approximations. 
We observe that the critical exponents that we have obtained by including next-to-leading order terms are closer to the best estimates in the literature than the critical exponents obtained at leading order and that the difference is compatible with the typical error bars in the Derivative Expansion approximation scheme. More generally, we have compared critical exponents obtained in the large-$N$ limit and near the upper critical dimension four, consistently demonstrating the quality of the approximations used.

We also determine the line $N_c^{+}(d)$ giving for each dimension the value of $N$ above which the transition is second order and below which it is first order. We obtain $N_c^{+}(d=3)=6.0(7)$ which is well above the physically relevant values $N=2$ or $N=3$. Complementarily, by integrating the RG flow, we specifically search  fixed points directly at $N=2$ and $N=3$ in $d=3$. 
It follows from these two analyzes that we do not find a fixed point in $d=3$ for $N=2$ or $N=3$ but, as previously obtained with the LPA and LPA', we find a region where the flow slows down, giving rise to a weakly first-order transition for a wide range of microscopic Hamiltonians. At the same time, we have been able to study regions of the parameter space (such as dimensions $d\simeq 2.4$) which complements studies performed with other methodologies (such as perturbation theory, Monte-Carlo simulations or Conformal Bootstrap).
In our view, these results reinforce the position that models with $O(N)\times O(2)$ symmetry exhibit a weakly first-order phase transition in $d=3$ for the experimentally interesting values of $N$, that is, $N=2$ and $N=3$.

We have also analyzed for the first time in the context of the FRG another type of second order phase transition called ``sinusoidal'' which near the upper critical dimension is observed for small values of $N<N_c^{-}(d)$. We have studied the curve $N_c^{-}(d)$, as well as another one for even smaller values of $N$ below which the theory exhibits a $O(2N)$ symmetry at the critical point. These studies have corroborated that the values of $N$ for which these transitions exist do not include $N=2$ or $N=3$. 

We do not want to end these conclusions without addressing possible future perspectives for the present work. First, ideally, it would be desirable to be able to include the full set of functions allowed by the symmetries at the next-to-leading order of the Derivative Expansion. Although direct analysis of the resulting equations seems out of reach in the short term, it is possible that by using the constraints arising from conformal invariance, such analysis will become affordable. We plan to address the corresponding study in the short term. Secondly, the experimental results show a set of exponents that seem to fall into two incompatible groups. While a first-order transition would not show universal (pseudo-)exponents and, therefore, there would be no reason why such exponents would have to be unique, the fact that they cluster into two sets is a relevant issue that 
is not very much discussed in the literature and that we intend to analyze also in the near future. Finally, the methods developed in the present work can be employed in other models such as the $q-$state Potts model, in which the authors has already made some contributions for $d \gtrsim 3$ dimensions. The techniques developed in the present work may allow to extend such analysis to lower dimensions which would allow to understand the passage from first to second order transitions also in that model.

\acknowledgements

We are very grateful to A. Stergiou for useful clarifications concerning
previous results on the large-$N$ limit of $O(N)\times O(2)$ models and M. Tissier and A. Codello for valuable
comments on the manuscript. C. S\'anchez and N. Wschebor thank for the
support of the Programa de Desarrollo de las Ciencias B\'asicas
(PEDECIBA). This work received the support of the French-Uruguayan
Institute of Physics project (IFU$\Phi$) and from the
grant of number FCE-1-2021-1-166479 of
the Agencia Nacional de Investigaci\'on e Innovaci\'on (Uruguay) and from the grant of number FVF/2021/160 from the Direcci\'on Nacional de Innovaci\'on, Ciencia y Tecnolog\'ia (Uruguay).

\appendix 

\section{The derivative expansion ansatz at $\mathcal{O}(\partial^2)$ for $O(N)\times O(2)$ models}
\label{apn:Ansatzes-DE2}

In this appendix we give the detailed deduction of the $\mathcal{O}(\partial^2)$ ansatz discussed in Sec.~\ref{DE2}. 
On top of the potential $U(\rho,\tau)$, all invariant terms with up to two derivatives of the fields $\vec{\phi}_1$ and $\vec{\phi}_2$ must be included. They are built out of $\partial_\mu \phi^i_a$ with $a \in \{1,2\}$ and $i \in \{1,2,\ldots,N\}$. 
The $O(\partial^2)$ ansatz can be written in a form that generalizes Eq.~\eqref{eq:DE2}
\begin{equation}\label{eq:DE2ansatz}
\Gamma[\vec\phi_1,\vec\phi_2]=\int_x \biggl\{U(\rho,\tau)+\frac{1}{2}{\cal Z}_{(i,a),(j,b)}\partial_{\mu}\phi_a^i\partial_{\mu}\phi_b^j\biggr\},
\end{equation}
 where ${\cal Z}_{(i,a),(j,b)}$ is a function of the components of $\vec\phi_1$ and $\vec\phi_2$ and where the index $k$ has been omitted.
Therefore, the problem consists in building explicitly the functions ${\cal Z}_{(i,a),(j,b)}$ in terms of the components of $\vec\phi_1$ and $\vec\phi_2$. Notice that all functions ${\cal Z}_{(i,a),(j,b)}$ evaluated in a constant field configuration can be defined from the second functional derivative of $\Gamma_k[\vec\phi_1,\vec\phi_2]$ evaluated in this configuration: 
 \begin{equation}
 \Gamma^{(2)}_{(i,a),(j,b)}(\vec\phi_1,\vec\phi_2)=\frac{\delta^2 \Gamma}{\delta\phi_a^i\delta\phi_b^j}
\end{equation}
because ${\cal Z}_{(i,a),(j,b)}$ is nothing but the $p^2$ term of $\Gamma^{(2)}_{(i,a),(j,b)}$ once this function is evaluated in a constant field and is Fourier transformed.

The $O(N)$-invariant terms are of two possible forms:
\begin{align}
 &A_{ab}=\partial_\mu \phi_a^i \partial_\mu \phi_b^i\nonumber\\
 &B_{abcd}=\varphi_a^i \varphi_b^j \partial_\mu \varphi_c^i \partial_\mu \varphi_d^j
\end{align}
where $A$ and $B$ are respectively  rank-two and rank-four $O(2)$ tensors. Obviously, $O(N)\times O(2)$ scalars involving two derivatives must be built out of either $A$ or $B$ contracted with $O(N)$ scalars that do not involve any derivative and that are respectively rank-two or rank-four $O(2)$ tensors. We have seen that all such tensors are obtained by tensor products of the identity $\delta_{ab}$ and  $S_{ab}=\phi^i_a \phi^i_b$. We find by inspection twelve invariant terms and ${\cal Z}_{(i,a),(j,b)}\partial_{\mu}\phi_a^i\partial_{\mu}\phi_b^j$ can be explicitly written
\begin{align}
&\frac{Z(\rho,\tau)}{2}\partial_\mu \phi_a^i \partial_\mu \phi_a^i
+\frac{Y(\rho,\tau)}{4}\phi_a^i \phi_b^j \partial_\mu \phi_a^i \partial_\mu \phi_b^j
\nonumber\\
&+W_1(\rho,\tau) S_{a b}\partial_\mu \phi_a^i  \partial_\mu \phi_b^i
+W_2(\rho,\tau) \phi_b^i \phi_a^j \partial_\mu \phi_a^i \partial_\mu \phi_b^j\nonumber\\
&+W_3(\rho,\tau) \phi_b^i \phi_b^j \partial_\mu \phi_a^i \partial_\mu \phi_a^j
+W_4(\rho,\tau) S_{a b}\phi_c^i \phi_c^j \partial_\mu \phi_a^i \partial_\mu \phi_b^j\nonumber\\
&+W_5(\rho,\tau) S_{a c}\phi_c^i \phi_b^j \partial_\mu \phi_a^i \partial_\mu \phi_b^j\nonumber\\
&+W_6(\rho,\tau)S_{a d}\phi_b^i \phi_d^j \partial_\mu \phi_a^i \partial_\mu \phi_b^j\nonumber\\
&+W_7(\rho,\tau)S_{a c}\phi_a^i \phi_c^j \partial_\mu \phi_b^i \partial_\mu \phi_b^j\nonumber\\
&+W_8(\rho,\tau)S_{a c}S_{b d}\phi_c^i \phi_d^j \partial_\mu \phi_a^i \partial_\mu \phi_b^j\nonumber\\
&+W_{9}(\rho,\tau) S_{a d}S_{b c}\phi_c^i \phi_d^j \partial_\mu \phi_a^i \partial_\mu \phi_b^j\nonumber\\
&+W_{10}(\rho,\tau) S_{a b}S_{c d}\phi_c^i \phi_d^j \partial_\mu \phi_a^i \partial_\mu \phi_b^j.
	\label{eqn:terms_ansatz_DE2_frust}
\end{align}
 However, as is the case for invariants without derivatives, not all invariant terms in  Eq.~(\ref{eqn:terms_ansatz_DE2_frust})  are independent. 
To show this, let us consider the structure of the matrix  ${\cal Z}_{(i,a),(j,b)}$ evaluated  in constant fields. As explained in Sec.~\ref{sec:LPA}, one can consistently choose a coordinate system where $\vec \phi_1$ and $\vec \phi_2$ take the form given by Eq.~(\ref{uniffield}). 
Furthermore, we can order the double indices in the following way:
\begin{align}
\Big((i,a)\Big)&=
\Big((1,1),(2,2),(2,1),(1,2),\nonumber\\
&(3,1),\dots,(N,1),(3,2),\dots,(N,2)\Big).
\end{align}
In this basis, the matrix ${\cal Z}_{(i,a),(j,b)}$ becomes block-diagonal: 
\begin{equation}
	\big(Z_k\big)=\left(\begin{array}{c |c |c |c}
		Z_1 & 0 & 0 & 0\\
		\hline
		0 & Z_2 & 0 & 0\\
		\hline
		0 &0  & D_1 & 0\\
		\hline
		0 & 0 & 0 & D_2\\
	\end{array}\right),
	\label{eqn:Gamma2strucfrust}
\end{equation} 
where $Z_1$ and $Z_2$ are symmetric $2 \times 2$ matrices, and the matrices $D_1$ and $D_2$ are proportional to the $(n-2)$-dimensional identity matrix. As a consequence, among the twelve terms in (\ref{eqn:terms_ansatz_DE2_frust}) at most eight are independent. To determine which of these terms are, we explicitly compute the non-zero components of ${\cal Z}_{(i,a),(j,b)}$:
\begin{align}
{\cal Z}_{(1,1),(1,1)}&=Z+2 \phi _1^2 \big(\frac{Y}{4}+W_1+W_2+W_3\big)\nonumber\\
   &+2 \phi _1^4 \big(W_3+W_5+W_6+W_7\big)\nonumber\\
   &+2 \phi _1^6 \big(W_8+W_{9}+W_{10}\big) \nonumber\\
{\cal Z}_{(1,1),(2,2)}&= \phi _1 \phi _2 \frac{Y}{2}+\phi _1 \phi _2 \big(\phi _1^2+\phi _2^2\big)  W_5+2 \phi _1^3 \phi _2^3 W_8 \nonumber\\
{\cal Z}_{(2,2),(2,2)}&=Z+2 \phi _2^2 \big(\frac{Y}{4}+W_1+W_2+W_3\big)\nonumber\\
&+2 \phi _2^4  \big(W_4+W_5+W_8+W_9\big)\nonumber\\
&+2 \phi _2^6 \big(W_8+W_{9}+W_{10}\big)\nonumber\\
{\cal Z}_{(2,1),(2,1)}&=Z+2 \phi_1^2 W_1+2 \phi_2^2 W_3+2\phi_1^2 \phi_2^2 W_4+2 \phi_2^4 W_7\nonumber\\
&+2 \phi_1^2 \phi_2^4 W_{10}\nonumber\\
{\cal Z}_{(2,1),(1,2)}&=2 \phi_1 \phi_2 W_2+\phi_1 \phi_2 \big(\phi_1^2+\phi_2^2\big)  W_6+2 \phi_1^3 \phi _2^3 W_{9}\nonumber\\
{\cal Z}_{(1,2),(1,2)}&=Z+2 \phi_2^2 W_1+2 \phi_1^2 W_3+2\phi_2^2 \phi_1^2 W_4+2 \phi_1^4 W_7\nonumber\\
&+2 \phi_2^2 \phi_1^4 W_{10}\nonumber\\
{\cal Z}_{(2,1),(2,1)}&=Z+2 \phi_1^2 W_1+2 \phi_2^2 W_3+2\phi_1^2 \phi_2^2 W_4+2 \phi_2^4 W_7\nonumber\\
&+2 \phi_1^2 \phi_2^4 W_{10}\nonumber\\
{\cal Z}_{(2,1),(1,2)}&=2 \phi_1 \phi_2 W_2+\phi_1 \phi_2 \big(\phi_1^2+\phi_2^2\big)  W_6+2 \phi_1^3 \phi _2^3 W_{9}\nonumber\\
{\cal Z}_{(1,2),(1,2)}&=Z+2 \phi_2^2 W_1+2 \phi_1^2 W_3+2\phi_2^2 \phi_1^2 W_4+2 \phi_1^4 W_7\nonumber\\
&+2 \phi_2^2 \phi_1^4 W_{10} \nonumber\\
{\cal Z}_{(i,1),(i,1)}&=Z+2 \phi_1^2 W_1\hspace{1cm}\text{for}\,i>2 \nonumber\\
{\cal Z}_{(i,2),(i,2)}&=Z+2 \phi_2^2 W_1\hspace{1cm}\text{for}\,i>2
\end{align}
where we omitted the $\rho$ and $\tau$ dependence of the various functions for notational symplicity. 
Solving this system of equations, we find  a set of eight independent functions leading to the  ansatz (\ref{eq:DE2ONO2}).

\section{Large$-N$ limit of $O(N)\times O(2)$ symmetric models within FRG}
\label{app:largeN}

The large$-N$ limit of $O(N)\times O(2)$ models is well understood using standard methods based on re-summation of perturbative diagrams. These diagrams are computed
around zero field configurations and we would like to understand how the large $N$ limit works in external field which is necessary for the study of the effective potential. 

We generalize the classical analysis of D'Attanasio and Morris \cite{DAttanasio:1997yph} for $O(N)$ models to the case of $O(N)\times O(2)$ symmetry. In the $O(N)$ models they found the general structure of the regularized effective action along the flow and we generalize this result below. We also solve the corresponding equations in the large $N$ limit.

\subsection{The structure of the effective action of $O(N)\times O(2)$ models in the $N\to\infty$ limit}

First, it is useful to observe that the standard microscopic Hamiltonian given by Eq.~(\ref{hamilitonian}) can be written in the following form
\begin{equation}
\label{hamiltonianbis}
H[\vec\varphi]=\int d^dx \Big\{ \frac{1}{2}\big(\partial_\mu\vec\varphi_{a}(x)\big)^2 + U(S)\Big\}.
\end{equation}
where $S$ is the rank-two $O(2)$ tensor with components $S_{ab}=\vec\varphi_a\cdot \vec\varphi_b$. Generalizing Ref.~\cite{DAttanasio:1997yph}, we now prove that the regularized effective action has the form
\begin{equation}
\label{GammalargeN}
\Gamma_k[\vec \phi]=\int d^dx \frac{1}{2}\big(\partial_\mu\vec\phi_{a}(x)\big)^2 + \hat \Gamma_k[S],
\end{equation} 
when $N$ is large.\footnote{The proof  presented here generalizes to arbitrary $O(N)\times O(M)$ models when $N$ is large, keeping $M$ fixed.} This is clearly the case in the initial condition of the RG flow since $\Gamma_\Lambda=H$. Following Ref.~\cite{DAttanasio:1997yph}, we assume that the effective action keeps the form (\ref{GammalargeN}) until an arbitrary scale $k_0$ at large $N$. Let us point out that this not true if $N$ is finite where, for example, the kinetic term acquires a running pre-factor $Z_k(\vec\phi)$ corresponding to a term different from those in Eq.~(\ref{GammalargeN}). Let us now prove that it stays true for $k=k_0-\delta k$ when $N$ is large. From Eq.~(\ref{GammalargeN}) we obtain
\begin{equation}
 \frac{\delta \Gamma_k}{\delta \phi_a^i(x)}=-\partial^2\phi_{a}^i(x)
 +2 \frac{\delta \hat\Gamma_k}{\delta S_{ac}(x)}\phi_{c}^i(x).
\end{equation}
To compute the flow equation of $\Gamma_k$ we need its second functional derivative. We evaluate it by keeping only the terms that contribute in the large $N$ limit, that is, those that are proportional to the $O(N)$ identity matrix and that are therefore the leading terms after tracing.  Let us note that this assumes that the effective action remains regular in the $N\to\infty$ limit for all $k\ge k_0$. Of course, if at a finite $k$ or, even, when $k$ goes to zero some couplings become singular, the present analysis needs to be corrected correspondingly (see, for example, \cite{Yabunaka:2023jlf}). The second functional derivative can be written as:
\begin{align}
 \frac{\delta^2 \Gamma_k}{\delta \phi_{a}^i(x)\delta \phi_{b}^i(x)}&=\delta_{ij}\Big(-\partial^2\big(\delta(x-y)\big)
 +2 \frac{\delta \hat\Gamma_k}{\delta S_{ab}(x)}\delta(x-y)\Big)\nonumber\\
 & + \mathrm{terms\,without\,}\delta_{ij}.
\end{align}
Inserting this expression in the flow equation (\ref{wettericheq}), one observes that the right hand side is a functional  of $S$ only. Given that we assume the form (\ref{GammalargeN}) until the scale $k_0$ and that this is also true for $\partial_k\Gamma_k$ at this scale, one concludes that the property remains true at $k=k_0- dk$. This, together with the validity of the property at the scale $k=\Lambda$ proves the property all along the flow.

It is interesting to point out an interpretation of the previous result. It implies (as for $O(N)$ models) that when $N\to\infty$, the regularized effective action $\Gamma_k$ is invariant  under {\it gauged} $O(N)$ transformations modulo the kinetic term that keeps its bare form. This means, for example, that the $Z(\vec\phi)$ function appearing in front of the kinetic term remains equal to one and, consequently, $\eta=0$.

\subsection{The large $N$ flow equation for the potential}

The exact flow equation for the potential reads
\begin{equation}\label{flowpot}
 \partial_t U_k(\rho,\tau)=\frac 1 2 \int_{q} \partial_t R_k(q) \mathrm{Tr}\Big[\Big(\Gamma^{(2)}(q;\vec\phi_a)+R_k\Big)^{-1}\Big],
\end{equation}
where the the right hand side, that involves the propagator must be calculated in a {\it constant field}. Now, given the property proved in the previous section:
\begin{equation}
 \Gamma_{ai,bj}^{(2)}(q;\vec\phi)=\delta_{ij}\big(q^2+2\partial_{S_{ab}}U_k\big)+ \mathrm{terms\,without\,}\delta_{ij}.
\end{equation}
To invert $\Gamma^{(2)}+R$, we rewrite it as a $2\times 2$ matrix and neglect the terms non proportional to $\delta_{ij}$
\begin{widetext}
\begin{equation}
\Gamma_{ai,bj}^{(2)}(q;\vec\phi)=\delta_{ij}
\left(\begin{array}{cc}
q^2+\partial_{\rho}U_k+2(\vec\varphi_1^2-\vec\varphi_2^2)\partial_\tau U_k &
4 \vec \varphi_1\cdot\vec \varphi_2 \partial_\tau U_k\\
4 \vec \varphi_1\cdot\vec \varphi_2 \partial_\tau U_k &
q^2+\partial_{\rho} U_k+2(\vec\varphi_2^2-\vec\varphi_1^2)\partial_\tau U_k \\
\end{array}\right).
\end{equation}
The propagator in a constant field then reads:
\begin{align}
G_{ai,bj}(q,\vec \phi)&=\frac{\delta_{ij}}{\big(q^2+R_k(q)+\partial_{\rho} U_k\big)^2-8\tau \big(\partial_\tau  U_k\big)^2}\nonumber\\
&\times\left(\begin{array}{cc}
q^2+R_k(q)+\partial_{\rho} U_k+2(\vec\varphi_2^2-\vec\varphi_1^2)\partial_\tau U_k &
-4 \vec \varphi_1\cdot\vec \varphi_2 \partial_\tau U_k\\
-4 \vec \varphi_1\cdot\vec \varphi_2 \partial_\tau U_k &
q^2+R_k(q)+\partial_{\rho}U_k+2(\vec\varphi_1^2-\vec\varphi_2^2)\partial_\tau U_k \\
\end{array}\right).
\end{align}
\end{widetext}
Substituting it into Eq.~\eqref{flowpot}, one then obtains for $N\to\infty$:
\begin{equation}
\label{LPA}
\partial_t U_k(\rho,\tau)=N\int_q \frac{\partial_t R_k(q)
\big(q^2+R_k(q)+\partial_{\rho}U_k\big)}
{\big(q^2+R_k(q)+\partial_{\rho}U_k\big)^2-8\tau \big(\partial_\tau U_k\big)^2}.
\end{equation}
As expected, this coincides with the LPA equation for the effective potential when $N$ is large.

\subsection{Solving the large-$N$ LPA equation}

Equation (\ref{LPA}) has many interesting properties. First, if one expands the effective potential in powers of $\tau$:
\begin{equation}
 U_k(\rho,\tau)=\sum_{n=0}^{\infty}U_n(\rho)\tau^n
\end{equation}
then it is easy to show that the flow equation for the function $U_n(\rho)$ only depends on the functions $U_m(\rho)$ with $m\le n$. This triangular property implies that solving finite subsets of flow equations for the $U_n(\rho)$ functions remains exact since they are decoupled from higher order functions in the $\tau$-expansion.

A second property is that these equations can be solved analytically by using the same method that in Refs.~\cite{DAttanasio:1997yph,Berges:2000ew,Blaizot:2005xy}. In particular, this implies that $U_0(\rho)$ is the same as in the $O(2N)$ model. This means that  $C_+$ has the same $U_0(\rho)$ function as the $O(2N)$ invariant Heisenberg fixed point (see, for example, \cite{Blaizot:2005xy}). Similarly, the tri-critical $C_-$ fixed point has a the same (null) $U_0(\rho)$ than the Gaussian fixed point.  Anisotropies only play a role for $U_n(\rho)$ for $n>0$. 

\bibliographystyle{apsrev4-2}
\bibliography{Frustrated}

\end{document}